\documentclass[prd,fontsize=8,usenatbib,nofootinbib,superscriptaddress,twocolumn]{revtex4-2}
\usepackage{graphicx}
\usepackage{amssymb,amsmath}
\usepackage{epstopdf}
\usepackage[usenames,dvipsnames]{color}
\usepackage{multirow}
\usepackage{hyperref}

\newcommand{\bxi}{\bar\xi}

\newcommand{\aapr}{Astron.~Astrophys.~Rev.}
\newcommand{\physrep}{Phys.~Rep.}
\newcommand{\mnras}{MNRAS}
\renewcommand{\apj}{ApJ}
\newcommand{\apjl}{ApJL}
\newcommand{\apjs}{ApJS}
\newcommand{\jcap}{JCAP}
\newcommand{\aap}{A\&A}

\begin{document}
\title{
High-resolution cosmological simulations of primordial dark matter clustering under long-range and fractional forces
}
\author{Derek Inman}\email{derek.inman@riken.jp}\affiliation{RIKEN Center for Interdisciplinary Theoretical and Mathematical Sciences (iTHEMS), Wako, Saitama 351-0198, Japan}

\begin{abstract}
    Long-range attractive fifth forces can lead to exponential instabilities in the early Universe.  For fermions with a Yukawa coupling to a sufficiently light scalar mediator, rapid oscillations of the scalar field can lead to a conservative force law with fractional behaviour on sufficiently large scales.  We study cosmological systems evolving under both this fractional potential and the Newtonian potential using high-resolution N-body simulations.  We find that, at the same mass scale, halos that form under the fractional potential are much more dense than those that from the Newtonian potential.  However, we also find that the perturbed scalar field may have large fluctuations once halo sizes become comparable to an effective Compton length, which will modify subsequent clustering and collapse.
\end{abstract}

\maketitle

\section{Introduction}

    Instabilities from long-range forces are well known in astrophysics, from the Jeans instability in self-gravitating systems \citep{bib:Binney1987} to various electromagnetic instabilities in plasmas \citep{bib:Marcowith2016}.  In cosmology, the observed clustering of matter, primarily composed of the unknown cold dark matter (CDM), arises due to the partial suppression of the gravitational instability by cosmic expansion \citep{bib:Lifshitz1946} and is a key piece of evidence for the $\Lambda$CDM model \citep{bib:Huterer2023}.  An important question is to what extent this dark matter can interact non-gravitationally, such as scattering with baryons \citep{bib:Dvorkin2014,bib:AliHaimoud2024}, couplings to dark energy \citep{bib:Amendola2000,bib:Farrar2004,bib:Simpson2010}, or through self-interactions \citep{bib:Tulin2018,bib:Adhikari2022}.  If such interactions are short-range, collisional behaviour can manifest as friction or heating if frequent \citep{bib:AliHaimoud2019,bib:Fischer2021,bib:Poulin2023}, or as rare pairwise scatterings which nonetheless can lead to dramatic effects like core collapse \citep{bib:Spergel2000,bib:Nadler2023}.  On the other hand, if they are long-range then the interacting dark matter may behave more like a collisionless gravitational \citep{bib:Archidiacono2022,bib:Bottaro2024} or plasma \citep{bib:Lasenby2020,bib:Cruz2023,bib:DeRocco2024} system with particles interacting collectively with a mean field.

    Attractive long-range forces can significantly accelerate structure formation leading to nonlinear fluctuations - halos - even in the radiation era \citep{bib:Savastano2019}.  A simple way to realize this is with Yukawa couplings between a scalar field (the force mediator) and fermion particles (the dark matter) \citep{bib:Flores2021}.  Such couplings occur in the standard model and are also well-motivated by asymmetric dark matter models \citep{bib:Zurek2014}, with a concrete model related to early halo formation being given in \citep{bib:Flores2023d}.  The formation of such halos in an otherwise nearly homogeneous early Universe can have numerous consequences including baryogenesis \citep{bib:Flores2023b}, gravitational waves \citep{bib:Flores2023a,bib:Wang2025}, and magnetogenesis \citep{bib:Durrer2023}.  Even if the fermions are not all the dark matter today, they may have played an important role in producing the CDM either by heating the plasma to re-introduce freeze-out at lower redshifts \citep{bib:Flores2023c} or collapsing into primordial black holes (PBH) \citep{bib:Amendola2018,bib:Flores2021,bib:Flores2023d}.  Since PBH can form much later in this scenario than the standard one, some constraints do not apply \citep{bib:Picker2023a} allowing such PBH to impact super-massive black hole formation \citep{bib:Lu2024a} or explain the gamma ray excess in the galactic centre \citep{bib:Picker2023b,bib:Korwar2025}.

    If the scalar field is massless, the force is effectively the same as the gravitational one on subhorizon scales; however, the fermions periodically become relativistic as they propagate through the scalar medium \citep{bib:Domenech2021}.  This situation can be avoided if a scalar mass is introduced, at the cost of having a finite range for the interaction \citep{bib:Domenech2023}.  That is, halos which are larger than the scalar Compton wavelength will not form from this force.  Curiously, however, Ref.~\citep{bib:Domenech2023} found that if the scalar potential is dominated by a quartic self-interaction, the potential is only partially screened as the effective scalar mass periodically oscillates to zero.  If the frequency is fast compared to the fermion dynamics, then the oscillation-averaged force is stable in comoving coordinates and long-range.  

    The formation of halos under such a screened force was demonstrated using N-body simulations in \citep{bib:Domenech2023}.  However, because the simulations evolved the oscillating system they were by necessity very low-resolution and the nonlinear regime could not be fully studied.  The goal of the present work is to implement the oscillation-averaged force law into high-resolution simulations, and to determine the extent that early Universe structure is sensitive to different attractive force laws.  \S~\ref{sec:linear} provides an overview of the long-range forces considered in this paper, and also serves as a brief review of Ref.~\citep{bib:Domenech2023} such that the present paper is self-contained.  \S~\ref{sec:quasilinear} describes our implementation of these forces using a high-resolution $P^3M$ force calculation and presents a comparison of the density statistics of fields evolved under the oscillation average potential to a more standard $1/r$ potential.  Lastly, \S~\ref{sec:nonlinear} discusses how the fractional force law begins to break down once the scalar field becomes sufficiently nonlinear.  

\section{Linear Potentials}
\label{sec:linear}

    In this section we show how the combination of higher-order, scalar field self-interactions with scalar-fermion Yukawa interactions can lead to long-range forces with fractional force laws.  Throughout this section we will assume that both the scalar and fermions can be linearized.  The cosmological case with quartic self-interactions (corresponding to $m=2$ below) was originally pointed out in Ref.~\citep{bib:Domenech2023} and the reader is referred there for a more thorough discussion, noting that the notation has been simplified here to reflect the oscillation-averaged limit.

\subsection{Force Classification}
\label{ssec:theory_force}
    Before discussing specific models, we provide in this section some basic definitions and discussion on long-range and fractional forces.  Specifically, we will consider conservative forces, $\vec{F}\propto-\vec{\nabla}\Phi$, where the potential $\Phi$ satisfies a linear elliptic equation,
    \begin{align}
        \hat{\mathcal{Y}}^{-1}\Phi\equiv(\nabla^2-\xi^{-2})\Phi = S \label{eq:poisson}
    \end{align}
    where $\hat{\mathcal{Y}}$ is a linear operator containing a Laplacian and a Debye-like screening parameter $\xi$, while $S$ is a source term, both of which can be time-dependent.  When $\xi=0$, this corresponds to the Newtonian or attractive Coulomb operator, $\hat{\mathcal{G}}$, which has the Green's function $\mathcal{G}$
    \begin{align}
        \mathcal{G}_k=-\frac{1}{k^2} \leftrightarrow \mathcal{G}_r=-\frac{1}{4\pi r} \label{eq:newtonian}
    \end{align}
    where subscript $k$ and $r$ indicate representations in Fourier space and real space respectively.  For $\xi>0$, we instead have the Yukawa interaction with Green's function $\mathcal{Y}$
    \begin{align}
        \mathcal{Y}_{k\xi}=-\frac{1}{k^2+\xi^{-2}} \leftrightarrow \mathcal{Y}_{r/\xi}=-\frac{\exp[-r/\xi]}{4\pi r}. \label{eq:yukawa}
    \end{align}
    The Yukawa Green's function can be related to the Newtonian one via $\mathcal{Y}_{k\xi}=\mathcal{A}_{k\xi}\mathcal{G}_k$ with 
    \begin{align}
    \mathcal{A}_{k\xi}\equiv\frac{1}{1+(k\xi)^{-2}}.
    \end{align}
    
    From a thermodynamics perspective (i.e.,~infinite particles at fixed density), forces are classified as long-range if their potentials decay slower than $-1/r^3$ as subsystems are then non-additive \citep{bib:Campa2009}.  In this work, we are more concerned with the formation of quasi-stationary states - in cosmology, halos - for which a more restrictive definition exists based on dynamics in the mean-field limit (i.e.,~infinite particles at fixed volume). Specifically, only a potential that decays slower than $-1/r^2$ is guaranteed to allow for halos which virialize but do not reach thermal equilibrium in finite time \citep{bib:Joyce2010}.  For either definition, the Newtonian interaction is an example of a long-range force whereas the Yukawa interaction is short-range.  On the other hand, if $\xi$ is much larger than the scale of the system (or a causal horizon), it is still reasonable to refer to such a force qualitatively as long-range \citep{bib:Campa2009}.  Collective effects become dominant over individual scatterings when the number of particles per Debye volume ($\sim\xi^3$) is large, which is called the collisionless limit \citep{bib:Campa2009,bib:Lasenby2020}.
    
\begin{figure}
        \includegraphics[width=0.45\textwidth]{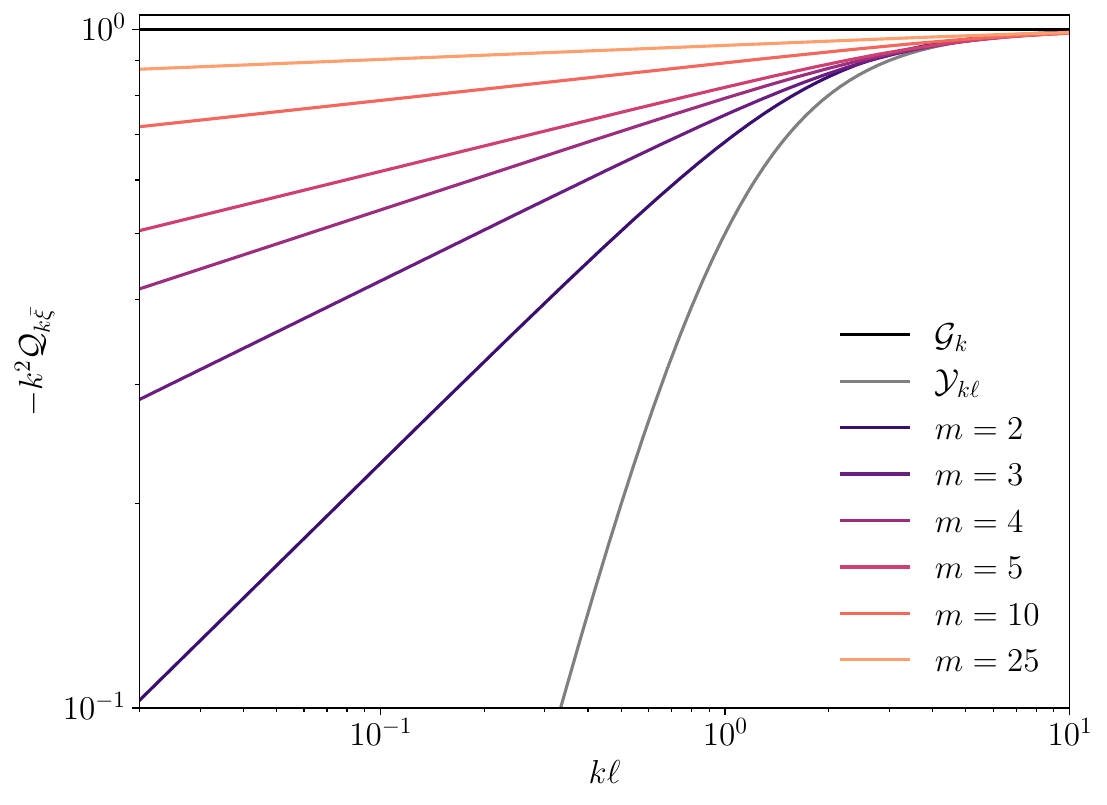}
        \caption{Scale-dependence of the Newtonian interaction ($\mathcal{G}$), the Yukawa interaction ($\mathcal{Y}$), and various fractional interactions ($\mathcal{Q}$).  The values of $\xi/\ell$ are chosen for each $m$ to match the fifth force arising in scalar-fermion systems.}
        \label{fig:oscGYQ}
    \end{figure}

    The operators $\hat{\mathcal{Q}}^{-1}$ modelling the scalar force that we will consider in the next section have Green's functions $\mathcal{Q}$ that are well approximated by
    \begin{widetext}
    \begin{align}
        \mathcal{Q}_{k\bxi}=A_{k\bar\xi}^{\frac{1}{4m-4}} \mathcal{G}_k \leftrightarrow \mathcal{Q}_{r/\bxi}=\left(\ _1F_2\left[ \frac{1}{4m-4};\frac{1}{2},1;\frac{r^2}{4\bxi^2}\right] - \frac{2}{\sqrt{\pi}}\frac{\Gamma\left[\frac{2m-1}{4m-4}\right]}{\Gamma\left[\frac{1}{4m-4}\right]}\frac{r}{\bxi}   \ _1F_2\left[\frac{2m-1}{4m-4};\frac{3}{2},\frac{3}{2};\frac{r^2}{4\bxi^2}\right]\right)\mathcal{G}_r \label{eq:qpotential}
    \end{align}
    \end{widetext}
    where $m\ge2$ is an integer and $\bxi$ is explicitly constant.  For $r\ll\bxi$ we have $\mathcal{Q}_{r/\bxi}\approx\mathcal{G}_r$ whereas for $r\gg\bxi$ we obtain $\mathcal{Q}_{r/\bxi}\propto-1/r^{(2m-1)/(2m-2)}$.  We compare the Newtonian, Yukawa and $\mathcal{Q}$ in Fig.~\ref{fig:oscGYQ} where we find that the latter lies between $\mathcal{G}$ and $\mathcal{Y}$.  We note that we have selected different $\bxi$ for each $m$ based on Eq.~\ref{eq:akxlt1} in the next section.

    The fastest this potential can decay is as $-1/r^{3/2}$ when $m=2$ and so it is also long-range by either definition.  However, this does not mean it is equivalent to other long-range forces.  The Newtonian interaction is particularly special as it is independent of scale and so the shell theorem holds \citep{bib:Martino2009}.  The potential in Eq.~\ref{eq:qpotential} falls off faster than the Newtonian potential, and so the relative effect of nearby particles will be larger.  Numerical simulations of a $-1/r^{3/2}$ potential have shown that this leads to faster relaxation than the Newtonian interaction for a finite number of particles \citep[see also \citep{bib:DiCintio2013}]{bib:Marcos2017}.  
    
    This potential can also be thought of as fractional as the long-range behaviour involves fractional powers of the Laplacian, i.e.,
    \begin{align}
        \hat{\mathcal{Q}}^{-1}\sim
        \begin{cases}
            -(-\nabla^2)^{\frac{4m-5}{4m-4}}\bxi^{-\frac{1}{2m-2}}  & {\rm large\ scales}\\
            -(-\nabla^2) & {\rm small\ scales}.
        \end{cases}
    \end{align}
    For a mathematical review on fractional Laplacians we refer the reader to \citep{bib:Lischke2020}, whereas a cosmology oriented discussion can be found in Appendix A of \citep{bib:Benetti2023}.  We note that in general the real and Fourier representations of the Green's function need not coincide for fractional Laplacians, but they do for infinite and periodic boundary conditions \citep{bib:Roncal2016,bib:Lischke2020}.

\subsection{Fractional Yukawa Forces}
\label{ssec:theory_fracY}
    Fractional force laws approximately of the form in Eq.~\ref{eq:qpotential} arise in non-expanding spacetimes as the effective force law mediated by a scalar field, $\phi$, with both a monomial self-interaction of strength $\lambda$ and a Yukawa interaction with fermions $\psi$ of strength $y$,
    \begin{align}
        V_{\rm eff}=\frac{\lambda}{2m}\phi^{2m} + y\phi n_\psi
    \end{align}
    where $n_\psi$ is the fermion particle density.  The scalar equation of motion is then given by the Klein-Gordon equation,
    \begin{align}
        \ddot\phi - \nabla^2\phi + \frac{\partial V_{\rm eff}}{\partial \phi}=0 \label{eq:scalareqn}
    \end{align}
    with $\partial V_{\rm eff}/\partial \phi=\lambda\phi^{2m-1}+y n_\psi$.  The case with $m=1$ corresponds to the standard Yukawa equation under which the fermions interact via the potential given by Eq.~\ref{eq:yukawa}, while $m=2$ is the cosmological case discussed in Ref.~\citep{bib:Domenech2023} and in \S\ref{ssec:theory_expU}.
    
    To show that this leads to fractional behaviour, let's now assume that the scalar field and fermions have a meaningful background about which they can be expanded, $\phi=\phi_0(t)+\phi_1(t,\vec{x})$ and $n_\psi=\bar n(1+\delta_\psi(t,\vec{x}))$.  The mean fermion density, $\bar n,$ is constant due to particle number conservation.  On the other hand, the mean scalar field evolves according to
    \begin{align}
        \ddot\phi_0+\lambda\phi_0^{2m-1}+y \bar n=0, \label{eq:phi0}
    \end{align}
    which can be constant, if $V_{\rm eff}$ is minimized, but can also change with time.

\begin{figure}
        \includegraphics[width=0.45\textwidth]{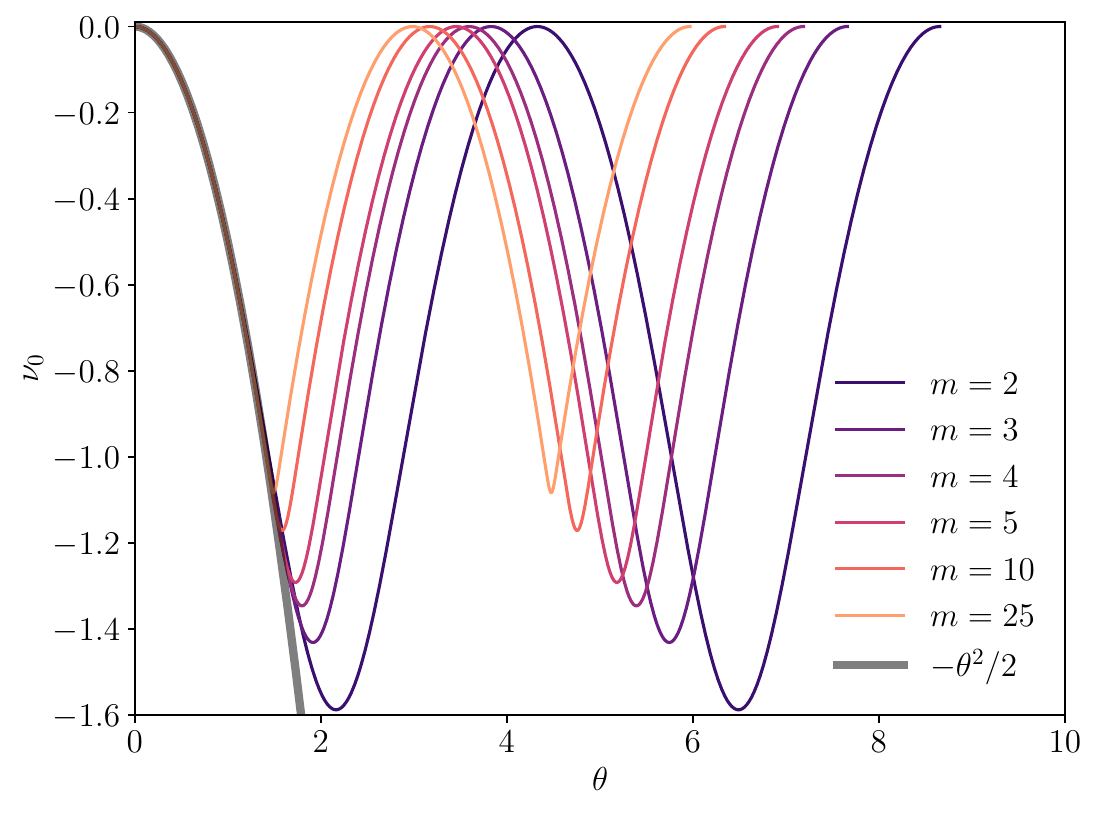}
        \caption{Background evolution of a scalar field for various scalar self-interactions $\propto\phi^{2m}$.  When the scalar field is large ($\nu_0\ll0$) it's potential is dominated by self-interactions ($\propto\nu_0^{2m}$), whereas when it is small ($\nu_0\sim0$) it is dominated by the Yukawa interactions and the scalar field behaves parabolically $\nu_0\approx-\theta^2/2$.}
        \label{fig:oscillations}
    \end{figure}

    We can simplify this equation by using dimensionless variables, $\phi_0(t)=\bar\phi \nu_0(\theta)$ with $\bar\phi=(y \bar n/\lambda)^{1/(2m-1)}$ and $\theta=\omega_\phi t$ where the generalized Compton frequency is $\omega_\phi^2=(\lambda(y \bar n)^{2m-2})^{1/(2m-1)}$, leading to the scale-free equation
    \begin{align}
        \frac{d^2\nu_0}{d\theta^2}+\nu_0^{2m-1}+1=0. \label{eq:nu0}
    \end{align}
    When the scalar field is small, $\nu_0\sim\nu_0'\sim0$, the Yukawa interactions dominate the self-interactions leading to $\nu_0\approx-\theta^2/2$.  When the self-interactions dominate, $\nu_0\ll0$, they instead provide a restoring force and return the scalar field to the origin.  We show this result for various monomial powers in Fig.~\ref{fig:oscillations}.

    The dynamical evolution of the perturbed scalar field is given by an inhomogeneous non-linear wave equation, 
    which is in general very difficult to solve but reduces in the non-relativistic limit to the Poisson-like equation,
    \begin{align}
        (\nabla^2-M^2_\phi)\phi_1=y \bar n \delta_\psi \label{eq:PoissonPhi1}
    \end{align}
    where $M^2_\phi=\lambda\left( (\phi_0+\phi_1)^{2m-1}-\phi_0^{2m-1} \right)/\phi_1$ is the effective mass of the scalar field.  If we now furthermore assume that the scalar field can be treated perturbatively then $M^2_\phi \phi_1\approx \partial^2V_{\rm eff}/\partial\phi^2=\lambda(2m-1)\phi_0^{2m-2}\phi_1=\ell^{-2} \nu_0^{2m-2}\phi_1$ with $\ell=(\sqrt{2m-1}\omega_\phi)^{-1}$ being a generalized Compton length scale.  Under these two approximations, the scalar equation becomes Eq.~\ref{eq:poisson} with $\xi/\ell=1/|\nu_0|^{m-1}$. 
    
    As pointed out in Ref.~\citep{bib:Domenech2023} for the quartic potential, $\nu_0\rightarrow0$ periodically, and so it is clear that the Yukawa potential $\mathcal{Y}_{k\xi}$ is at least occasionally long-range for $m\ge2$.  Furthermore, the timescale of these oscillations is $\sim\omega_\phi^{-1}$ which can be very small for systems with sufficiently high $\bar n$.  In this case, the oscillations may be significantly faster than the fermion dynamics and we can perform an oscillation average of $\mathcal{Y}_{k\xi}$, which we define as $\mathcal{Q}_{k\bxi}=\langle \mathcal{Y}_{k\xi}\rangle=\langle \mathcal{A}_{k\xi} \rangle \mathcal{G}_k$ with,
    \begin{align}
        \langle \mathcal{A}_{k\xi} \rangle = \frac{1}{P} \int_0^P \frac{1}{1+\nu_0^{2m-2}/(k\ell)^2} d\theta \label{eq:oscavgF}
    \end{align}
    where $P$ is the period of oscillations.  For $m=1$ this is simply the constant Yukawa interaction, $\langle\mathcal{A}_{k\xi}\rangle=\mathcal{A}_{k\ell}$.  For $m\ge2$ the oscillation average can be computed numerically; however, we can extract the long-range behaviour, $k\ell\ll1$, analytically by noting that in this regime only $\nu_0\sim-\theta^2/2\sim0$ contributes significantly to the integral
    \begin{align}
        \langle \mathcal{A}_{k\xi} \rangle_{k\ell\ll1} &\approx \frac{1}{P} \int_0^\infty \frac{1}{1+(-\theta^2/2)^{2m-2}/(k\ell)^2} d\theta \nonumber \\
        &=\frac{2\sqrt{2}}{P} \Gamma\left[\frac{4m-5}{4m-4}
        \right]\Gamma\left[\frac{4m-3}{4m-4}\right] (k\ell)^{\frac{1}{2m-2}} \nonumber \\ 
        &\equiv(k\bxi)^{\frac{1}{2m-2}}\label{eq:akxlt1}
    \end{align}

\begin{figure}
        \includegraphics[width=0.45\textwidth]{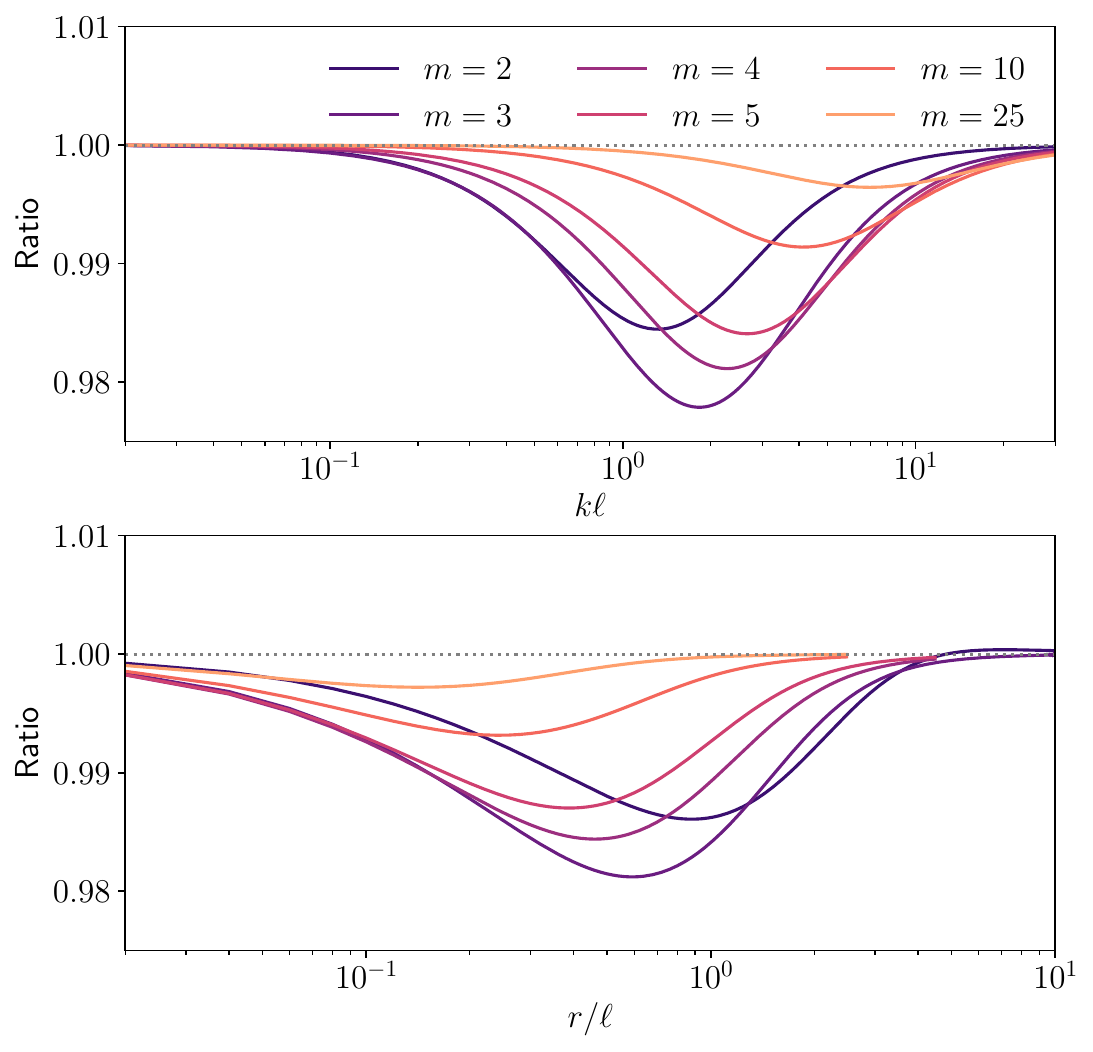}
        \caption{Ratio of approximate potential $\mathcal{Q}$ in Eq.~\ref{eq:qpotential} to the numerical evaluation of Eq.~\ref{eq:oscavgF}.  The top panel shows the Fourier space calculation whereas the lower panel shows the corresponding real space calculation.  For large $m$ the real space curves are truncated due to numerical oscillations when evaluating $_1F_2$ with large arguments.}
        \label{fig:oscFacc}
    \end{figure}

    Ref.~\citep{bib:Domenech2023} found that $\langle\mathcal{A}_{k\xi}\rangle\approx \mathcal{A}_{k\bxi}^{1/4}$ is a very good approximation for $m=2$ even with an approximate relation for $\bxi/\ell$.  We generalize this result to higher $m$ in Eq.~\ref{eq:qpotential}: $\langle \mathcal{A}_{k\xi} \rangle \approx \mathcal{A}_{k\bxi}^{1/(4m-4)}$ with $\bxi$ related to $\ell$ based on Eq.~\ref{eq:akxlt1}.  We demonstrate the accuracy of the approximation for a variety of $m$ in Fig.~\ref{fig:oscFacc}, finding that the maximum deviation is around $\sim2\%$.  
    We further note that Ref.~\citep{bib:Domenech2023} found an exact solution to Eq.~\ref{eq:nu0} with $m=2$ in terms of elliptic Jacobi equations which yields for the period $P=(4\sqrt{\pi}/2^{1/6})(\Gamma[7/6]/\Gamma[2/3])$ and so $\bxi\approx \ell/1.89735$.

\subsection{Validity for Dark Matter}
\label{ssec:theory_expU}
    
    In an expanding Universe, where background and perturbatations can be properly modelled, the Klein-Gordon equation additionally depends on the scalefactor, $a$,
    \begin{align}
        \phi''+2\frac{a'}{a}\phi'-\nabla^2\phi+a^2\frac{\partial V_{\rm eff}}{\partial \phi} = 0
    \end{align}
    where primes denote derivatives with respect to conformal time, $\eta$ defined by $a d\eta=dt$.  Following Ref.~\citep{bib:Domenech2023}, we can simplify this substantially via the transformation $\phi=(a_i/a)\varphi$ after which we obtain:
    \begin{align}
        \varphi''-\frac{a''}{a}\varphi-\nabla^2\varphi+\frac{a^3}{a_i}\left( \lambda \frac{a_i^{2m-1}}{a^{2m-1}}\varphi^{2m-1} + y n_\psi\right)=0. \label{eq:varscalareqn}
    \end{align}
    In the radiation dominated Universe we have that $a=H_i a_i^2 \eta$ and so $a''/a\equiv0$.  If we furthermore assume non-relativistic fermions $n_\psi=n_i a_i^3/a^3(1+\delta_\psi)$ we see that if $m=2$ then Eq.~\ref{eq:varscalareqn} is equivalent to Eq.~\ref{eq:scalareqn} with the associations $t\rightarrow\eta$, $\lambda\rightarrow\lambda a_i^2$ and $y \bar n \rightarrow y n_i a_i^2$.  Using the same steps as before to get to Eq.~\ref{eq:nu0}, we have then that $\bar\phi=a_i\bar\varphi=(y n_i a_i^3/\lambda)^{1/3}$ and $\omega_\varphi^2=(\lambda(y n_i a_i^3)^2)^{1/3}$ from which we see that $\omega_\varphi^2$ is sourced by the comoving fermion number density $n_i a_i^3={\rm const}$.
    The corresponding cosmological Compton length scale $\ell$, which we now associate to $\omega_\varphi^2=(3\ell)^{-2}$, is therefore constant in comoving coordinates.

    Now let us consider the dynamics of the fermions in this same epoch, again following Ref.~\citep{bib:Domenech2023}.  At the background level, the fermions have an effective mass $m_{\rm eff}=m_\psi + y \phi_0$ which decays to $m_\psi$ provided $y\bar\varphi/m_\psi\ll1$ and so the effect can be neglected.  The linear equations of motion are then given by,
    \begin{align}
        \eta^2\delta_\psi''+\eta\delta_\psi'=\frac{s}{4}\frac{k^2}{k^2+M_\varphi^2}\delta_\psi \approx \frac{s}{4}\langle A_{k\xi}\rangle \delta_\psi
        \label{eq:lin_delta_psi}
    \end{align}
    where
    \begin{align}
        s=12 y^2\frac{M_{\rm pl}^2}{m_\psi^2}\frac{m_\psi n_i}{\rho_i} \frac{a}{a_i}=s_i \frac{a}{a_i} \label{eq:s_i}
    \end{align}
    and $\rho_i$ is the background radiation density with $\rho_i=3 H_i^2 M_{\rm pl}^2$ and $M_{\rm pl}^2=(8\pi G)^{-1}$.
    The growing solution for the fermion density contrast is then
    \begin{align}
        \delta_\psi(s,k)=\sqrt{\Delta_0(k)} \frac{I_0(\sqrt{\langle A_{k\xi}\rangle s})}{I_0(\sqrt{\langle A_{k\xi}\rangle s_0})}
    \end{align}
    where we can take $s_0\rightarrow0$ with nonzero $\Delta_0$.
    
    The density contrast on scales $k\bxi\gg1$ begins to grow significantly after $s=s_i\eta/\eta_i\gtrsim1$, suggesting the appropriate fermion time scale is $\omega_\psi^{-1}=\eta_i/s_i$.  As discussed in Ref.~\citep{bib:Domenech2023}, the oscillation average procedure should be appropriate when the typical scalar oscillation timescale is much smaller
    \begin{align}
        \frac{\omega_\varphi^{-1}}{\omega_\psi^{-1}} = 12 \left(\frac{1}{\lambda^{1/6}y^{1/3}}\right)\left(\frac{y^2}{m_\psi^2/M_{\rm pl}^2}\right)\left(\frac{m_\psi n_i}{\rho_i}\frac{H_i}{n_i^{1/3}}\right) \ll 1. \label{eq:oscavg_validity}
    \end{align}
    The first term in the parenthesis corresponds to the mean field limit, which should be valid provided there are a large number of particles within a characteristic volume,
    \begin{align}
        n_i a_i^3 \left(\sqrt{3}\ell\right)^3 = \frac{1}{\sqrt{\lambda}y} \gg 1, \label{eq:meanfield_validity}
    \end{align}
    or, in other words, we need fairly small couplings.  The second term corresponds to the relative strength of the fifth force to gravity.  Like other forces, we expect that that the scalar force is stronger than gravity implying
    \begin{align}
        1\gtrsim y\gg m_\psi/M_{\rm pl}. \label{eq:yukawa_validity}
    \end{align}
    So far, the requirements of Eq.~\ref{eq:meanfield_validity} and~\ref{eq:yukawa_validity} oppose Eq.~\ref{eq:oscavg_validity}.  However, the third term in Eq.~\ref{eq:oscavg_validity} can be very small.  For instance, if the fermions are stable and make up all the dark matter then $m_\psi n_i/\rho_i\sim a_i/a_{\rm eq}$ and $(a_i/a_{\rm eq})(H_i/n_i^{1/3})\sim10^{-25}(m_\psi/{\rm GeV})^{1/3}$.  Note that this quantity is independent of the arbitrary parameter $a_i$ and so Eq.~\ref{eq:oscavg_validity} is either always satisfied or never satisfied in the radiation era.  
    
    The next section describes numerical simulations in the limit set by Eq.~\ref{eq:oscavg_validity}.  While this allows us to model nonlinear density fields, it remains in the same simplified system discussed here and in Ref.~\citep{bib:Domenech2023}: in a radiation-dominated Universe  with relevant scales being subhorizon and appropriate thermodynamic considerations satisfied (see also \citep{bib:Domenech2021}); with gravitational accelerations which are negligible;\footnote{The omitted gravitational term is $(3/2)(a/a_i)(m_\psi n_i / \rho_i)\delta_\psi$ on the right-hand side of Eq.~\ref{eq:lin_delta_psi}.  Requiring this to be much smaller than $(s/4)\langle \mathcal{A}_{k\xi}\rangle\delta_\psi$ and assuming Eq.~\ref{eq:yukawa_validity} then requires $k\bxi\gg1/(\sqrt{2}y M_{\rm pl}/m_\psi)^4$.} using growing-mode Gaussian initial conditions rather than a consistent inflationary treatment, (one such example being \citep{bib:Flores2023d}) and assuming negligible quantum corrections to the quartic potential (see, e.g.,~\citep{bib:Blinov2018,bib:CarrilloGonzalez2021} for discussions in other contexts.)

\section{Quasilinear Simulations}
\label{sec:quasilinear}

    The cosmic fermion density field becomes nonlinear exponentially fast, and so studying its' dynamics requires nonlinear modelling.  Initially, however, the scalar field fluctuations remain small and can be modelled by the linear Poisson-like equations sourced by the nonlinear fermion density.  Ref.~\citep{bib:Domenech2023} previously solved this system using N-body simulations that had very short time-steps in order to resolve the scalar oscillations controlling the effective fermion mass, $m_{\rm eff}$, and the effective force length scale, $\bxi$.  Unfortunately, these requirements necessitated a very low-force resolution (a particle mesh calculation), even for parameters that only mildly satisfied the oscillation-average approximation and which are far away from cosmologically relevant values.  The high level of force softening further made it difficult to study the nonlinear density field deep inside  halos.  Our goal in this section then is to replace this calculation with one assuming the oscillation average procedure, and then use a higher resolution force calculation.  For the fractional force, we make use of Eq.~\ref{eq:qpotential} with $m=2$ throughout.  
    
    We have also implemented simulations using the Newtonian potential, which corresponds to simply replacing $\mathcal{Q}$ with $\mathcal{G}$ in the subsequent equations.  On physical considerations, a truly massless scalar field leads to relativistic fermions \citep{bib:Domenech2021}; however, we could consider it to represent the quadratic case in \citep{bib:Domenech2023} where $a m_\phi^{-1}$ is simply large compared to the simulation volume so that $\mathcal{Y}\approx\mathcal{G}$.  Similarly, we could consider this as the dynamics of the $m\ge2$ potential on wavenumbers much larger than $\bxi^{-1}$ as again $\mathcal{Q}\approx\mathcal{G}$.  To be explicit, these simulations are to model Yukawa-type interactions and not the gravitational accelerations, which are always neglected.

\subsection{Equations of Motion}
    Our first goal is to find the correct oscillation-averaged equations of motion.  The general form of Hamilton's equations are given by \citep{bib:Domenech2023}:
    \begin{align}
        \frac{d\vec{x}}{d\eta} = \frac{\vec{p}}{a m_{\rm eff}}\ ;\ \frac{d\vec{p}}{d\eta} = - a_i y \vec{\nabla} \varphi_1
    \end{align}
    where $\vec{x}$ is the comoving particle coordinate and $\vec{p}=m_{\rm eff} \vec{v}$ the conjugate momentum.  We now assume that all non-oscillating quantities ($\vec{x}$, $\vec{p}$, $s$, $\delta_\psi$) are constant over a period of scalar oscillations, $\Delta\eta=P/\omega_\psi$.  Based on the arguments and calculations in Ref.~\citep{bib:Domenech2023} we furthermore assume that $m_{\rm eff}$ decays rapidly to $m_\psi$ and that its' oscillations can also be neglected.  Taking the oscillation average of the equations of motion then leads to
    \begin{align}
        \frac{\vec{x}(\eta+\Delta\eta)-\vec{x}(\eta)}{\Delta\eta} &= \frac{1}{\Delta\eta}\int_\eta^{\eta+\Delta\eta} \frac{\vec{p}}{a m_\psi} d\eta'\\
        &\approx \frac{\vec{v}}{a} \label{eq:osc_avg_x}
    \end{align}
    and
    \begin{align}
       m_\psi \frac{\vec{v}(\eta+\Delta\eta)-\vec{v}(\eta)}{\Delta\eta} =-\frac{a_i y}{\Delta\eta}\int_\eta^{\eta+\Delta\eta} 
         \vec{\nabla}\varphi_1(\eta',x(\eta'))d\eta' \label{eq:osc_avg_v}
    \end{align}
    When $\Delta\eta$ is small, the first equation is simply a definition of the derivative.  The second, however, defines the oscillation averaged acceleration if we assume that the particle trajectory over a period is effectively constant, $\vec{x}(\eta')\approx\vec{x}$.  Assuming $\Delta\eta$ is sufficiently small, we therefore replace Hamilton's equations with
    \begin{align}
        \frac{d\vec{x}}{d\tau}=\vec{v}\ ;\ \frac{d\vec{v}}{d\tau} = -\nabla \Phi
    \end{align}
    where $a d\tau=d\eta\gg\Delta\eta$ is the superconformal time \citep{bib:Martel1998} and $\Phi$ is the oscillation average of $a (a_i y/m_\psi) \varphi_1=(y/m_\psi)a^2\phi_1$ satisfying the fractional equation
    \begin{align}
        (H_i a_i^2)^{-2}\hat{\mathcal{Q}}^{-1}\Phi = \frac{s}{4}\delta_\psi. \label{eq:Qpoisson}
    \end{align}
    In practice, we can also compute additional corrections to these equations that are also linear in the potential $\Phi$.  We provide these in Appendix~\ref{app:expansion} and show that the non-decaying term is suppressed by Eq.~\ref{eq:oscavg_validity}.

    To solve these equations numerically, we use dimensionless simulation units and a drift-kick-drift leapfrog algorithm as in Ref.~\citep{bib:Domenech2023}.  We also use comparable timestep criteria, keeping $dt=d\log s\le0.01$, $dt\le \log[1+0.25/s]\rightarrow ds<0.25$, $dt\le r_{\rm min}/{\rm max}[v_p]$, and $dt\le \sqrt{0.04\ r_{\rm min}/{\rm max}[a_p]}$.  Crucially, we no longer need to limit the timestep based on the oscillating scalar field as we are now using oscillation-averaged equations.

\subsection{P$^3$M Force Calculation}
    When computing forces with high resolution, it is common to split the calculation into long-range and short-range components which are evaluated using different techniques.  Here, we use a Particle-Particle Particle-Mesh (P$^3$M) force calculation where we evaluate the long-range force spectrally ($\mathcal{Q}\rightarrow\mathcal{Q}_{k\bxi}$) and the short-range force directly ($\mathcal{Q}\rightarrow\mathcal{Q}_{r/\bxi}$).

    To compute the P$^3$M force, we need to first split the force into the PM and PP portions, as well as provide a softening scheme for the latter.  In general, these are fairly standard procedures used in large-scale structure for gravitational forces \citep{bib:Hockney1988,bib:HarnoisDeraps2013,bib:Angulo2022} and our goal here is to adapt them to the scale-dependent scalar force.

\subsubsection{PM Force}
    To obtain the grid-based PM acceleration, we first compute the density field $\delta_\psi$ by interpolating the N-body particles to a cubical grid using the Cloud-In-Cell (CIC) interpolation scheme.  The density field is then Fourier transformed and multiplied with the long-range part of the Green's function, which we define via a Gaussian splitting kernel, $\tilde{\mathcal{Q}}_{k\bxi}=\exp\left[-(k r_s)^2\right]\mathcal{Q}_{k\bxi}$.  Even for gravitational forces, there are several ways of implementing the Fourier space kernel \citep{bib:Feng2016}.  We opt for a modified version of kernel (iv) in \citep[and related to that of \citep{bib:Springel2021}]{bib:Feng2016},
    \begin{align}
        \vec{\nabla}\tilde{\mathcal{Q}}\rightarrow i\vec{D}_k\tilde{\mathcal{Q}}_{k\bxi} W_{\rm CIC}^{-2}
    \end{align}
    where $W_{\rm CIC}$ is the CIC convolution kernel, $\vec{D}_k$ corresponds to the finite difference gradient operator in Fourier space and we have replaced the gravitational kernel with $\mathcal{Q}_{k\bxi}$.  The field is then transformed back into Fourier space, multiplied by $s/4$ to yield the acceleration, and then re-interpolated back to the particle position using the same CIC scheme in order to conserve momentum.  

\subsubsection{PP Force}
    We next consider the pairwise PP force between particles that are a distance $r$ apart.  In Fourier space, the short-range part of the Green's function is simply $(1-\exp\left[-(k r_s)^2\right])\mathcal{Q}_{k\bxi}$ but in real-space there is a non-trivial convolution of $\mathcal{Q}_{r/\bxi}$.  To simplify the calculation, we can set the splitting scale to be substantially less than the Compton length, $r_s\ll\bxi$ such that $\mathcal{Q}\approx\mathcal{G}$.  The convolution with the Newtonian potential is exact and leads to a multiplicative factor in real space as well, $\tilde{\mathcal{G}}_{r}={\rm erfc}\left[r/(2r_s)\right]\mathcal{G}_{r}$ \citep{bib:Angulo2022}.  When computing the short-range scalar force, we assume this same multiplicative factor can be used as well; however, because the convolution is no longer exact, it is ambiguous in real space whether the approximation $\mathcal{G}\rightarrow\mathcal{Q}$ should be made for the potential kernel or the acceleration kernel.  We opt for the latter, and, as we are assuming $r\ll\bxi$, we Taylor expand the acceleration Green's function
    \begin{align}
        -4\pi r^2 \mathcal{Q}_{r/\bxi\ll1}'&\approx 1 - \frac{1}{8}\left(\frac{r}{\bxi}\right)^2\nonumber\\
        &+\frac{4}{9\sqrt{\pi}}\frac{\Gamma[\frac{7}{4}]}{\Gamma[\frac{1}{4}]}\left(\frac{r}{\bxi}\right)^3-\frac{5}{256}\left(\frac{r}{\bxi}\right)^4
    \end{align}
    such that the pairwise acceleration between two particles is given by
    \begin{align}
        \vec{a}_{\rm PP}=-\frac{\vec{r}}{4\pi r^3}&(-4\pi r^2 \mathcal{Q}_{r/\bxi\ll1}')\frac{s}{4}\frac{(H_ia_i^2)^2}{n_i}\nonumber\\&\left({\rm erfc}\left[\frac{r}{2 r_s}\right]+\frac{r}{\sqrt{\pi}r_s}\exp\left[-\left(\frac{r}{2 r_s}\right)^2\right]\right).
    \end{align}
    For softening, we assume the particles are a uniform sphere such that the force law is further transformed below a scale $r_{\rm min}$ \citep{bib:Dyer1993} 
    \begin{align}
        \frac{1}{4\pi r^3}\rightarrow\frac{1}{4\pi r_{\rm min}^3}\left(8-9\frac{r}{r_{\rm min}}+2\left(\frac{r}{r_{\rm min}}\right)^3\right).
    \end{align}
    In practice, we tabulate $|\vec{a}_{\rm PP}|/r$ and then later interpolate to find the acceleration.  As such, in the future we could remove some of the approximations we have made, e.g.,~by numerically performing the convolution with a Gaussian or by using the numerical evaluation of Eq.~\ref{eq:oscavgF}.  Our numerical implementation of the calculation follows the description in Ref.~\citep{bib:HarnoisDeraps2013}, utilizing a linked-list and looping over grid cells without repeating symmetric calculations to optimize speed. 

\subsubsection{Accuracy}

\begin{figure}
        \includegraphics[width=0.45\textwidth]{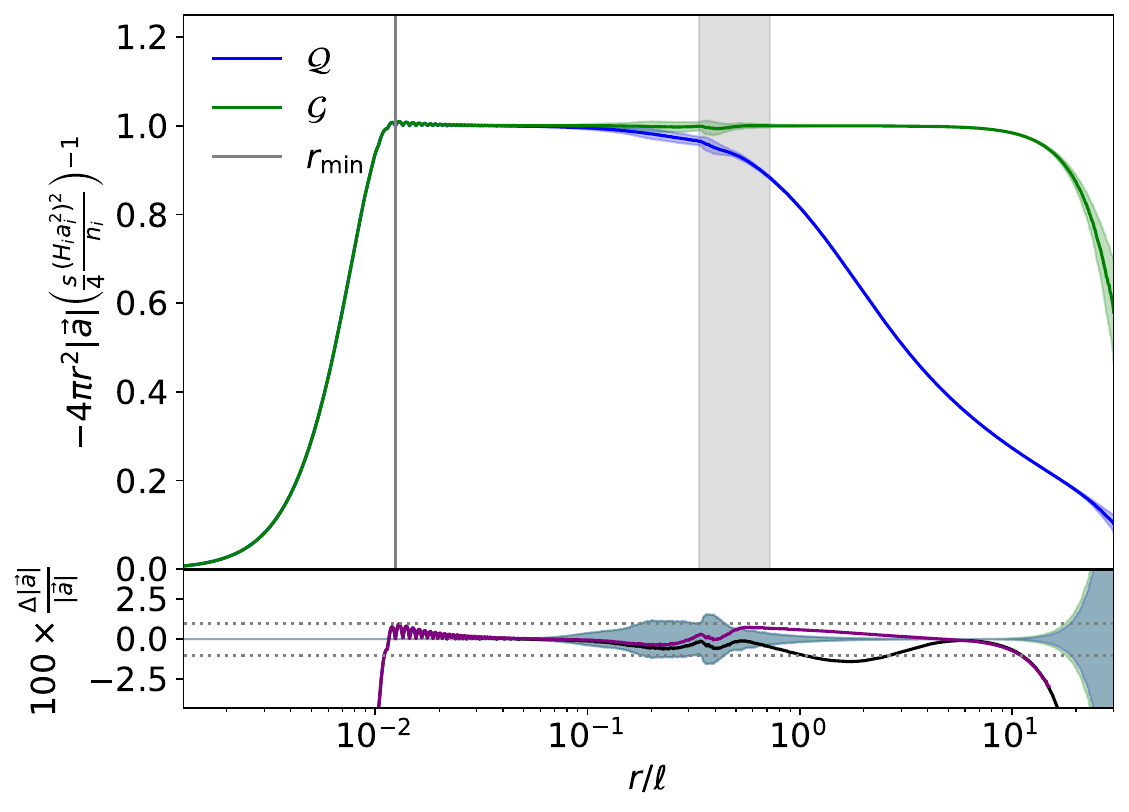}
        \caption{(Top) P$^3$M force calculation for an inverse square law (green) and the oscillation averaged scalar force (blue).  Solid lines show the mean of 1000 repetitions, while shaded regions show the $1\sigma$ standard deviation.  The vertical grey line shows the softening length, while the shaded grey region indicates where the grid geometry affects whether the PP force is calculated.  (Bottom)  Shaded regions show the standard deviation without the mean.  The solid curves show the fluctuation between the mean force and the approximate force based on Eq.~\ref{eq:qpotential} (purple) and an exact oscillation averaged force calculation (black).  Both theoretical and numerical errors are around $\sim 1\%$ (shown as dotted horizontal lines).}
        \label{fig:facc}
    \end{figure}

    So far, we have focused on the overall procedure; here we describe our specific implementation and accuracy tests.  We first specify the characteristic fifth force length scale in grid cells via $\ell/L=n_\ell/n_c$ where $L$ is the simulation volume and $n_c$ is the number of grid cells in the PM computation.  Based on the tests in Ref.~\citep{bib:Domenech2023} we choose $n_\ell=12$, and use grids of size $n_c=768$.  For the force splitting, we use a value of $r_s=1.15$ and compute the PP force in a cubic volume up to 4 grid cells away.  For the softening length, we use $r_{\rm min}=0.15$, approximately a tenth of the interparticle distance.  To quantify accuracy we use a similar force test as in Ref.~\citep{bib:Domenech2023}: we set one massive particle at a random location, and then compute the acceleration of 999 other massless particles with different $r$ and oriented randomly.  We then repeat this process 1000 times in order to obtain variances.  We show the result of this calculation in Fig.~\ref{fig:facc}.  We find the maximum variance occurs around the force splitting scale and is typically no more than $\sim1\%$.  The largest errors correspond to the region between $4$ and $5\sqrt{3}$ grid cells which is where particles may or may not have a PP force depending on their orientation with respect to the grid.  In the bottom panel we show the relative error between our numerical calculation of the force and the forces corresponding to the approximate Eq.~\ref{eq:qpotential} (i.e.~$\propto\partial\mathcal{Q}_{r/\bxi}/\partial r$) and the exact Eq.~\ref{eq:oscavgF} (i.e.,~$\propto\langle \partial\mathcal{Y}_{r\xi}/\partial r\rangle$), finding that they have comparable percent-level errors.

    In addition, a particles location relative to the grid can induce force artifacts that can build up over time.  A common way to reduce such errors is to uniformly shift particle positions so that they do not repeatedly have the same force error \citep{bib:HarnoisDeraps2013,bib:Springel2021}.  We therefore add a random shift up to 5 grid cells per dimension to the particle positions at the beginning of each timestep and substract the same random offset at the end of the timestep.

\subsection{Initial Conditions}
    \label{ssec:ic}
    We assume that fluctuations are initially Gaussian with a power law initial spectrum
    \begin{align}
        \Delta^2_0=\left(\frac{k}{k_0}\right)^{2(n_0-1)} \langle\delta_0^2\rangle_V
    \end{align}
    where $_V$ indicates a volume average (rather than a periodic average).  We set the initial variance as $\langle\delta_0^2\rangle_V=10^{-6}$, the pivot scale to be $k_0\ell=4/\Delta_{\rm cr}^{1/3}$ with $\Delta_{\rm cr}=200$, and the spectral index as $n_0=1$ (flat, similar to Ref.~\citep{bib:Domenech2023}) or $n_0=2$ (blue-tilted).  Simulations are not started at s=0, but rather at some later time $s_i=10^{-2}$.  For a physical model we should set $s_i$ based on Eq.~\ref{eq:s_i}, but in practice we treat it as a free parameter and choose $s_i=10^{-2}$.  A similar consideration holds for $k_0$ when $n_0\ne1$, although in this case it concerns where $\Delta_0^2\sim1$ relative to $\ell^{-1}$.
       
    For our comparison simulation using $\mathcal{G}$, the $n_0=1$ case cannot be simulated correctly as the largest scales will be dominated by the periodic boundary conditions.  As an alternative, we add additional scale dependence in the initial conditions such that the linear $\Delta_\psi^2$ at a given final scalefactor $s_f$ is equivalent to that of the $\mathcal{Q}$ simulation.  To do this, we introduce an inverse transfer function $I_0( \sqrt{-k^2\mathcal{Q}_{k\bxi}s_f} )/I_0(s_f)$ and can make use of the approximation $I_0(x\gg1)\approx \exp[x](1/\sqrt{2\pi x}+1/8/\sqrt{2\pi x^3}+9/128/\sqrt{2\pi x^5})$.  We therefore have three initial power spectra for four simulations.  
    
    We now need to express $\delta_\psi(s_i)$ in terms of particles, which is typically done through the use of the Zel'dovich approximation $\vec{\nabla}\cdot\vec{\Psi}=\delta_\psi$, with $\vec{\Psi}$ being the displacement field \citep{bib:Zeldovich1970}.  Here we find a conceptual difference between simulations in the matter era and the radiation era.  In a matter era the growth rate is $\propto a$ and so it is sensible to assume that there are no initial particle displacements, or $\delta_0\rightarrow 0$ as $a\rightarrow0$.  However, in a radiation era this is not the case as $\delta_0$ remains finite as $s_0\rightarrow 0$.  In other words, there is a non-zero primordial displacement $\vec{\nabla}\cdot\vec{\Psi}_0=\delta_\psi(s_0)$ and whether to move the particles by $\vec{\Psi}$ or $\vec{\Psi}-\vec{\Psi}_0$ is unclear.
    
\begin{figure}
        \includegraphics[width=0.45\textwidth]{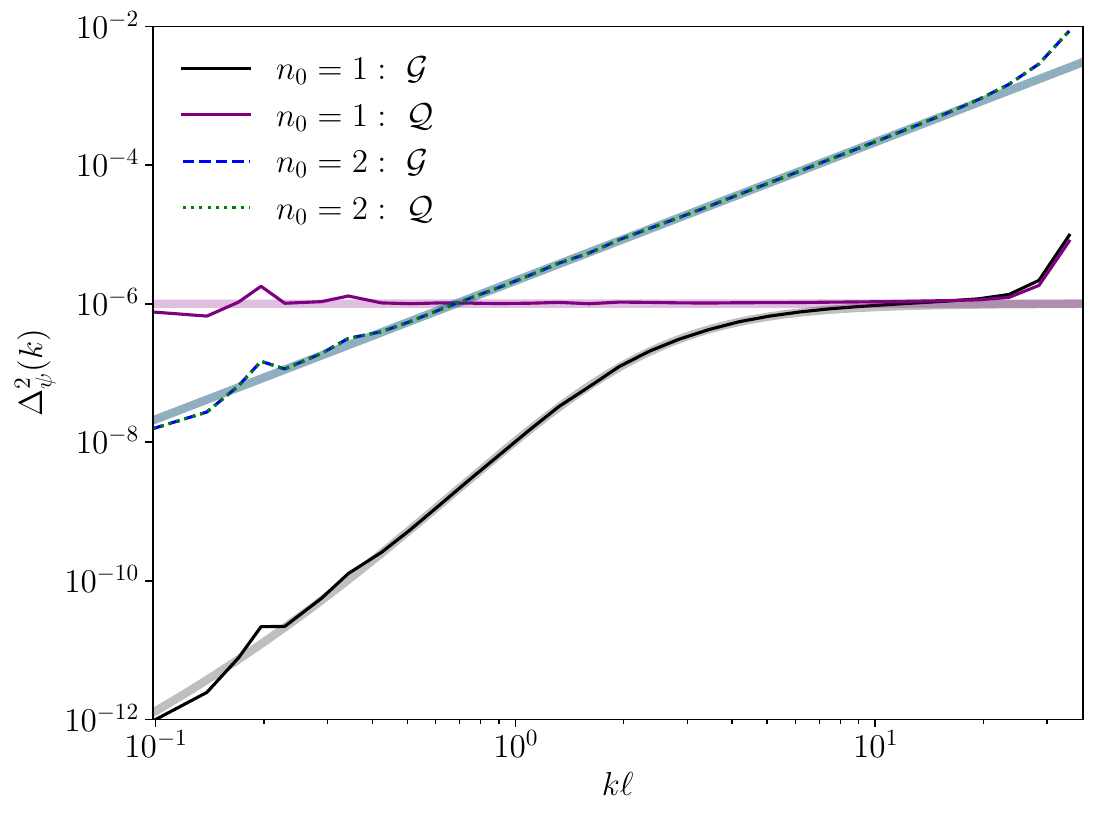}
        \caption{Initial power spectra at $s_i=10^{-2}$.  While the two $n_0=2$ simulations have the same initial power spectrum, the $n_0=1$ simulations are different at $s_i$ but have the same linear power at a later redshift.}
        \label{fig:pki}
    \end{figure}

    We consider this ambiguity in more detail in Appendix~\ref{app:primdisp}.  In practice we do not find much difference based on how we intialize particles, and so we have opted to use a standard approach using $\vec{\Psi}$ (see Eq.~\ref{eq:displacement1}), primarily because it allows us to straightforwardly have fewer particles than grid cells.  Specifically, we use a body-centered cubic lattice with $N_p=2\times(n_c/2)^3$ total particles \citep{bib:Joyce2005,bib:Marcos2008}.  For reference, we show the initial power spectra at $s_i$ in Fig.~\ref{fig:pki}.

\subsection{Correlation Function}
    For studying the density field, we make use of the real-space two-point correlation function $\Xi(r)$.\footnote{We use capital $\Xi(r)$ rather than the more common $\xi(r)$ to avoid confusion with the length scale $\xi$.}  Commonly, this calculation can be done either by Fourier transforming the power spectrum \citep{bib:Karamanis2021},
    \begin{align}
        \Xi(r,s) = \int_0^\infty \Delta^2(k,s) j_0(k r) \frac{dk}{k} \label{eq:pkxi}
    \end{align}
    or in a pairwise manner between particles or galaxies \citep{bib:Sinha2020}.  Here, we instead compute it based off density fields using the grid-based method described in \citep{bib:Inman2017}: given an initial density grid, we compute correlations between cells a distance of $\sqrt{1}$, $\sqrt{2}$, and $\sqrt{3}$ cells apart; then we reduce to a smaller grid by a factor of $2$ or $3$ via averaging; we repeat this process until the reduced grid size is just one cell.
    
    So far, we have not specified how to obtain the density grid as the best choice of particle interpolation schemes is not obvious.  One option is to use the CIC interpolation scheme, which we also used in the PM force calculation.  However, this scheme leads to an auto-correlation for individual particles which is somewhat undesirable.  Furthermore, reducing a grid by averaging is not equivalent to interpolating directly to the smaller grid size.  The simpler Nearest-Grid-Point (NGP) method has neither of these issues, but has a less smooth density field with potentially larger interpolation artifacts. 
    
    We provide some detailed tests of the particle interpolation scheme in Appendix \ref{appsub:xipartinterp}.  In general, we have opted to use the NGP scheme for particle interpolation based on the considerations above, but have confirmed that the choice does not affect our results significantly.  In order to probe down to the softening scale, we use a grid size a factor of four larger than in the P$^3$M calculation (e.g.,~3072 grid cells per dimension).  Further convergence tests of our results with respect to simulation parameters are provided in Appendix \ref{appsub:sftest}.

\begin{figure}
        \includegraphics[width=0.45\textwidth]{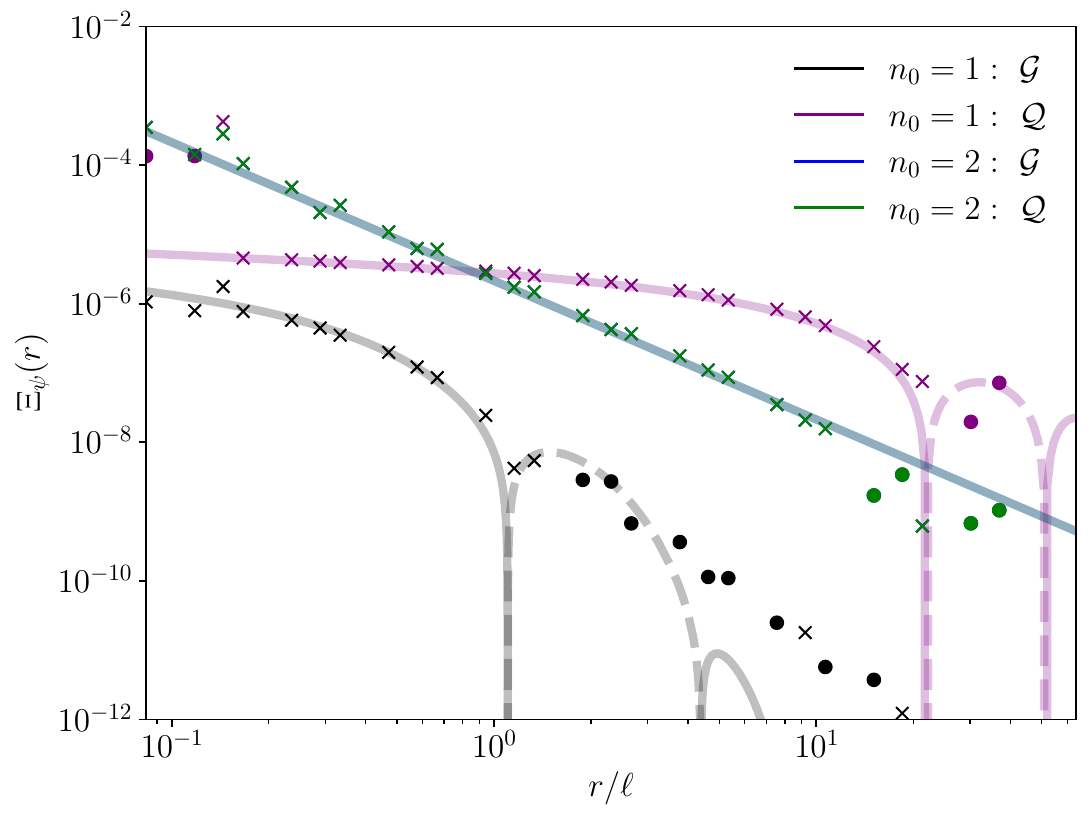}
        \caption{Two point correlation function of the initial conditions.  Crosses and dots indicate the positive and negative values of the correlation function measured from initial particle configuration.  Solid and dashed lines correspond to theoretical predictions at $s_0$.}
        \label{fig:xii}
    \end{figure}

    One exception to the above is during the initial linear evolution.  Here, the CIC scheme reproduces the smooth density field well (with Poisson noise being cancelled by the lattice initialization) while NGP has large interpolation errors when assigned to the grid.  We therefore use the same size grid as the PM force, the CIC interpolation scheme, and have also shifted particles to a cell vertex by subtracting $0.5$ cells from each position.  We show the correlation functions of the initial conditions described above in Fig.~\ref{fig:xii}.  To compare our numerical calculation with linear theory, we have done slightly different calculations for each simulation.  In all cases, since the power spectrum at $s_i$ is very similar to $s_0$, we simply show the $\Xi(r)$ at $s_0$ curves as these can be integrated more straightforwardly for power law initial conditions.  The most difficult calculation is the flat initial conditions ($n_0=1,\ \mathcal{Q}$) as power from all scales contributes equally. We therefore use Eq.~\ref{eq:pkxi} with $\Delta_0^2$ set to $0$ outside $k_{\rm min}=2\pi/L$ and $k_{\rm max}=\pi n_c/L$ where $L=\ell n_c/n_\ell$.  For the scale-dependent $n_0=1$ case with $\mathcal{G}$, we instead perform the calculation numerically using ${\rm hankl}$ \citep{bib:Karamanis2021} with $k_{\rm min}$ reduced and $k_{\rm max}$ increased by a factor of $10$ to reduce ringing.  Lastly, for the $n_0=2$ cases we use $\Xi(r)=\langle \delta_0^2 \rangle_V/(k_0\ell)^2 \left(\frac{r}{\ell}\right)^{-2}$.     
    
    In general we find good agreement between the grid-based calculation and the analytic estimates based on power spectra, except at very large scales.  Matching such scales in real space precisely is nontrivial (see \citep{bib:Pen1997}) and it could be beneficial to re-examine our somewhat standard setup in future calculations.

\subsection{Simulation Results}
    \label{ssec:results}
    In this section, we present our results comparing halo formation under the fractional $\mathcal{Q}$ potential to that of the Newtonian potential $\mathcal{G}$.  In general, it is not obvious how to best make this comparison.  For instance, if we run simulations with the same initial conditions for the same amount of time then the halos that form under $\mathcal{G}$ will be substantially bigger than under $\mathcal{Q}$.  We therefore instead opt to compare the simulations at the same characteristic halo size.  We define this as in Ref.~\citep{bib:Domenech2023}, occurring when the linear matter power spectrum is unity, $\Delta^2(s,k_{\rm nl})=1$, and relating to the typical halo mass by
    \begin{align}
        M_{\rm hl}=\frac{4}{3}\pi (m_\psi n_ia_i^3) \Delta_{\rm cr} R_{\rm hl}^3 = \frac{4}{3}\pi (m_\psi n_i a_i^3) k_{\rm nl}^{-3}
    \end{align}
    where $\Delta_{\rm cr}\equiv200$ is the mean overdensity of a halo.  As discussed in \S~\ref{ssec:ic} we will further consider simulations with the same initial conditions, i.e., the same $\delta_\psi(s_0)$ for $n_0=2$, and those with the same linear theory final conditions, i.e., the same $\delta_\psi(s_f)$ for $n_0=1$.  Table~\ref{tab:hlradius} shows the scalefactors corresponding to a few characteristic halo radii for each simulation setup.  
    
\begin{figure}
        \includegraphics[width=0.45\textwidth]{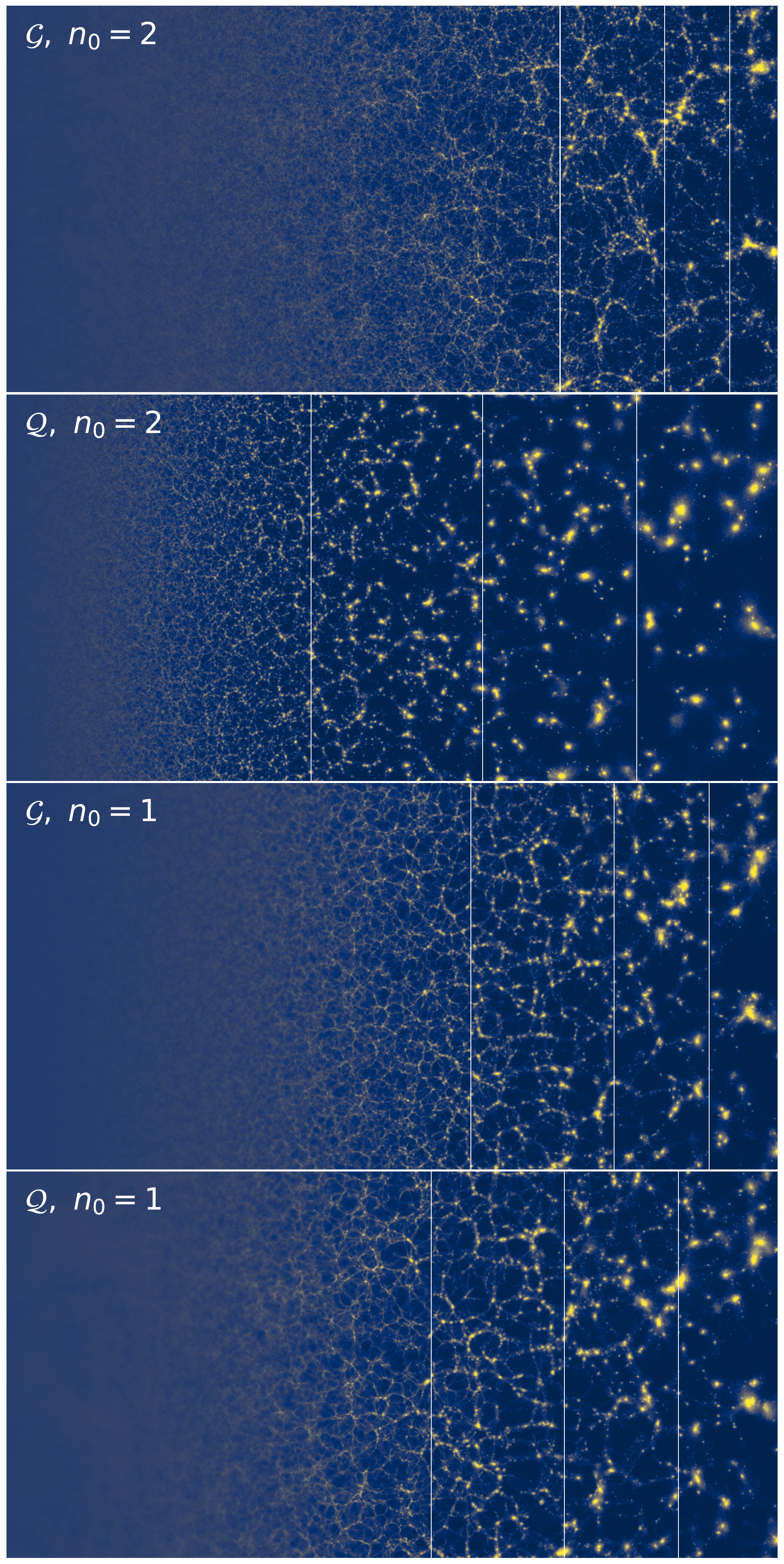}
        \caption{Visualization of spacetime evolution of the four simulations listed in Table~\ref{tab:hlradius} where the left edge corresponds to the same redshift $s_i=10^{-2}$, the right edge corresponds to when the typical halo radius is $R_{\rm hl}/\ell=1/2$ and other locations are equally separated in scalefactor relative to the particular simulation's final redshift.  Vertical white lines show when each simulation has $R_{\rm hl}/\ell=1/8,\ 2/8,$and $3/8$ for comparison.}
        \label{fig:slice4}
    \end{figure}

    We show a visualization of the spacetime evolution for each of the four simulations in Fig.~\ref{fig:slice4}.  Focusing on the $n_0=2$ cases, the simulation with the Newtonian potential (top panel) resembles the more familiar large-scale structure, having filaments connecting a web of halos, even at the final redshift.  Given that this simulation is scale-free (both in initial conditions and potential interaction), we can expect this to continue through larger scales until an effective Compton length $\ell$ is reached.  The simulation run with the fractional potential (upper middle panel) represents such a case and we see that once scales comparable to $\ell^{-1}$ become nonlinear the structure has collapsed into just halos, as was noticed in Ref.~\citep{bib:Domenech2023}.  Alternatively, we can interpret the $\mathcal{G}$ simulation not as an early phase of the $\mathcal{Q}$ one, but rather as a direct comparison of how different potentials lead to different evolution.  In this interpretation, comparing the top panel to the three other cases indicates that it is the scale-dependence - either of the force law or the initial conditions - that determines the structure properties.

    \begin{table}
    \begin{tabular}{|c||c|c|c|c|}
        \hline
        &\multicolumn{4}{c|}{Simulations} \\
        \cline{2-5}
        & \multicolumn{2}{c|}{Matched $s_0$} & \multicolumn{2}{|c|}{Matched $s_f$}\\
        \cline{2-5}
        $n_0$& \multicolumn{2}{c|}{$2$} & \multicolumn{2}{|c|}{$1$}\\
        \hline\hline
        & \multicolumn{4}{|c|}{Scalefactor $s$}\\
        \cline{2-5}
        $R_{\rm hl}/\ell$ & $\mathcal{Q}$ & $\mathcal{G}$ & $\mathcal{Q}$ & $\mathcal{G}$\\
        \hline
        $1/8$ & 87.2  & 66.7  & 103.7 & 113.4 \\
        $2/8$ & 136.2 & 79.3 & 136.2 & 148.3 \\
        $3/8$ & 180.3 & 87.1 & 164.0 & 171.6 \\
        $4/8$ & 220.6  & 92.9 & 188.3 & 188.3 \\
        \hline
    \end{tabular}
    \caption{Evolution of halo radius as a function of scalefactor for each of our four simulations.  When initial conditions are matched we use the blue-tilted power spectrum for both $\mathcal{Q}$ and $\mathcal{G}$ force laws.  When final conditions are matched we use a flat power spectrum for $\mathcal{Q}$ and a scale-dependent power spectrum for $\mathcal{Q}$.}
    \label{tab:hlradius}
    \end{table}

\begin{figure}
        \includegraphics[width=0.45\textwidth]{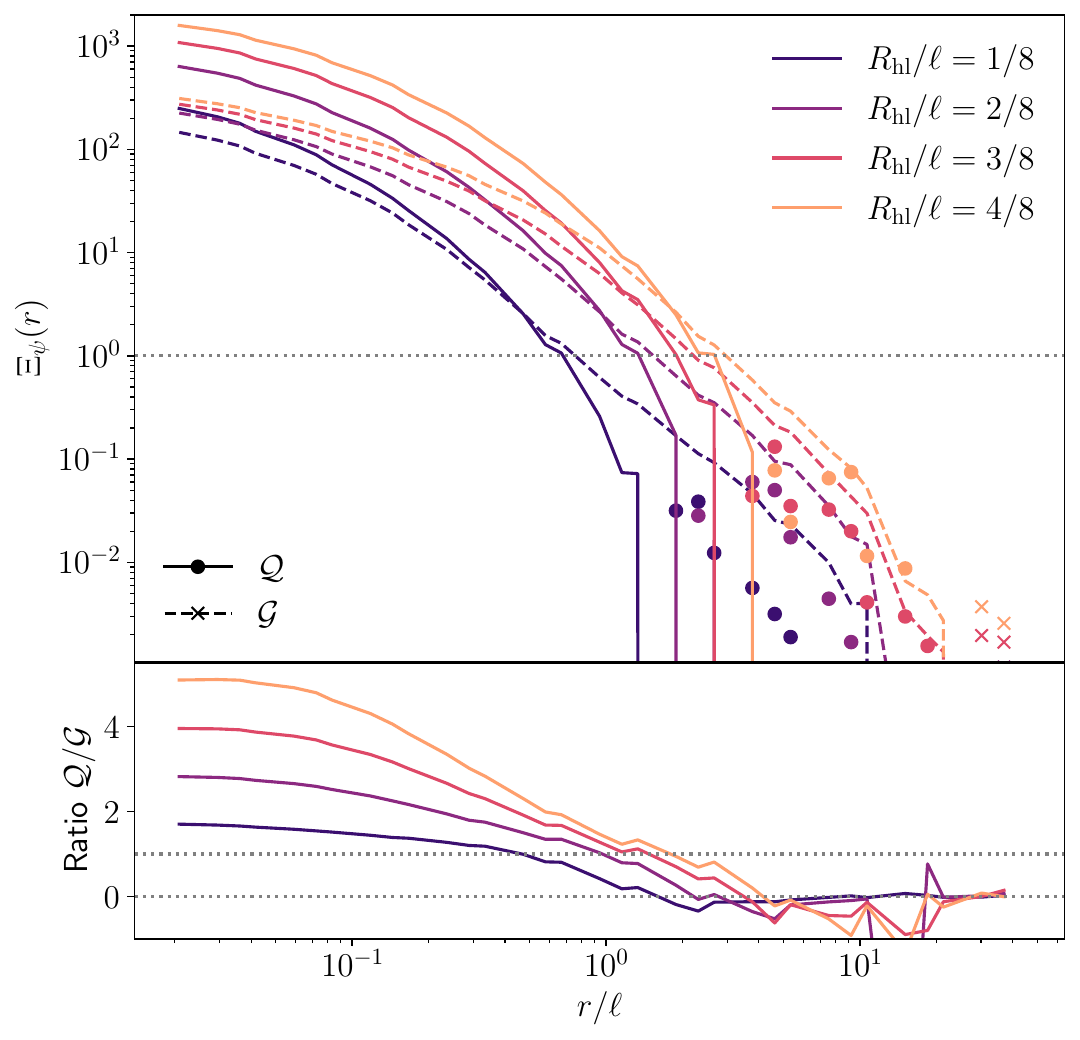}
        \caption{The real-space correlation function assuming matched initial conditions.  The correlation function is systematically larger for the oscillation-averaged potential ($\mathcal{Q}$) than the Newtonian one ($\mathcal{G}$).  In the upper panel, negative values of the correlation function are shown via points instead of curves.}
        \label{fig:xiz_n2}
    \end{figure}

    To be more quantitative, we now consider the real-space correlation function.  We start by comparing the matched initial conditions simulations (the top two panels of Fig.~\ref{fig:slice4}) and show $\Xi_\psi(r)$ in Fig.~\ref{fig:xiz_n2} at various $R_{\rm hl}$.  We find that the correlation function on small-scales, related to typical halo densities, is substantially larger for simulations evolved under the $\mathcal{Q}$ potential.  The reason for this enhancement is not entirely obvious as the typical halo size is less than the force screening length and so the interior dynamics are always in the regime where $\mathcal{Q}_{r/\ell\ll1}\approx\mathcal{G}$.  In idealized gravity-based secondary-infall calculations, particle orbits become time-independent after relaxation, as shells of matter spend most of their time away from the center of the halo, and so the central mass does not build up with accretion \citep{bib:Bertschinger1985}.  This result relies on self-similarity however, and the $\mathcal{Q}$ potential explicitly breaks this as large halos are accreting from a region of size $R_{\rm nl}=k_{\rm nl}^{-1}=R_{\rm hl}\Delta_{\rm cr}^{1/3}$ which is greater than $\ell$ for $R_{\rm hl}/\ell\sim1.4/8$.  In other words, both the time between halo mergers and the in-falling halo trajectories can be different between $\mathcal{Q}$ and $\mathcal{G}$ forces.

\begin{figure}
        \includegraphics[width=0.45\textwidth]{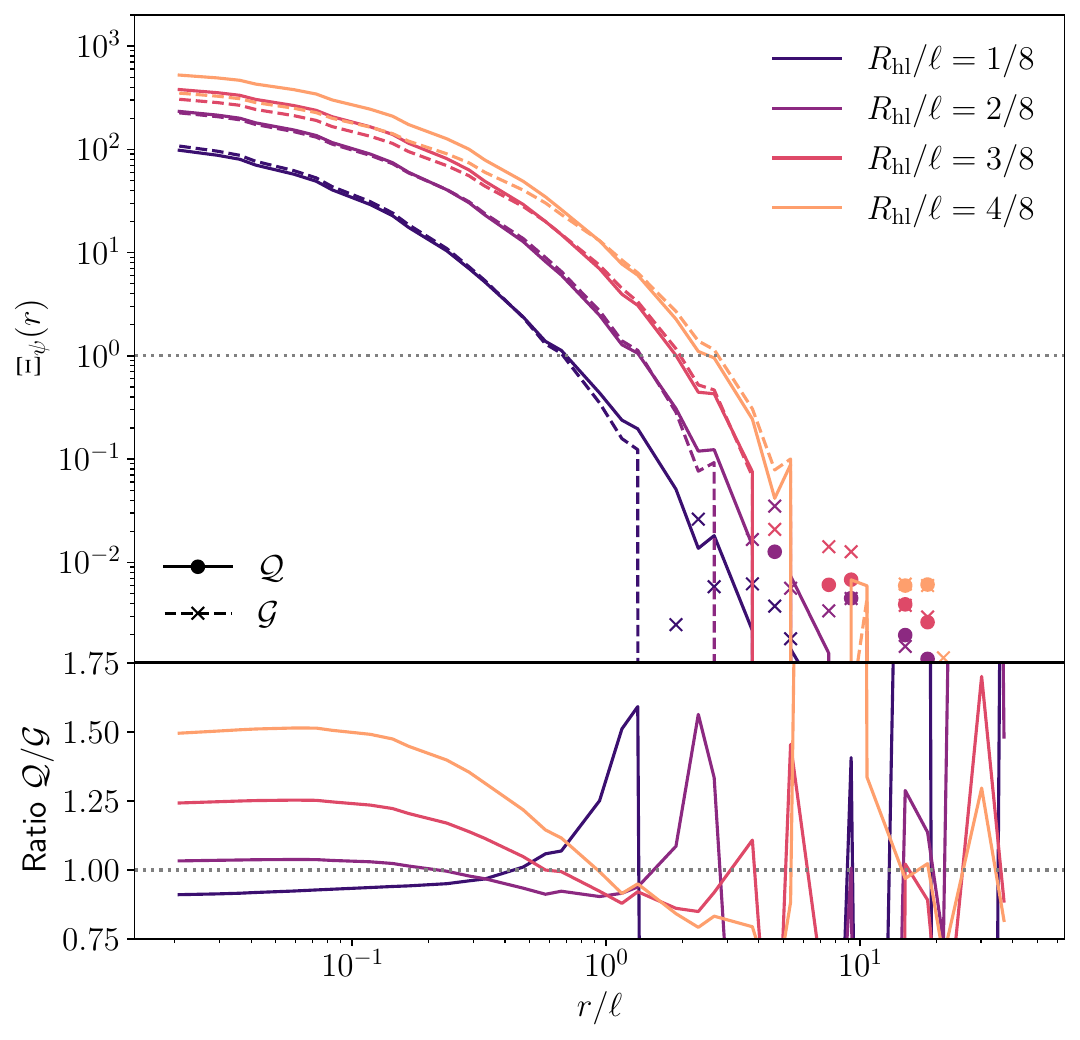}
        \caption{The same as Fig.~\ref{fig:xiz_n2} except for the simulations with matched (linear) final conditions.  Halos are still systematically more dense with the oscillation averaged potential, but not as substantially as when initial conditions are matched.}
        \label{fig:xiz_n1}
    \end{figure}

    To further test this interpretation, we can consider our simulations with matched final conditions which have much more similar evolution histories.  Fig.~\ref{fig:xiz_n1} shows the correlation functions with matched final conditions and indeed the densities are much more similar.  Nonetheless, there is still a residual enhancement at late times in the $\mathcal{Q}$ simulation indicating that simply changing the initial conditions cannot uniformly mimic a change in the force field.  Since the differences between these two simulations are small, it is important to consider whether they could be numerical artifacts.  We provide a number of convergence tests in Appendix~\ref{app:convergence} and find that the largest source of error appears to be in the calculation of the correlation function itself, up to $\sim10\%$.  Given that we are looking at ratios, we expect such errors to at least partially cancel; however, to be strict we can only conclude that the excess observed in the bottom panel of Fig.~\ref{fig:xiz_n1} are robust for $R_{\rm hl}\ell=3/8$ and larger.  Additionally, we find that the $\mathcal{Q}$ simulation with $n_0=2$ may have more numerical systematics as we observe worse convergence (a few percent, instead of sub-percent) when changing particle number than in the other cases.  We show this in more detail in Appendix~\ref{appsub:numericalparams}, where we also show the accuracy of our calculation on resolved scales within the halo (roughly, $10^{-1}\ell\lesssim r \lesssim \ell$).

\begin{figure}
        \includegraphics[width=0.45\textwidth]{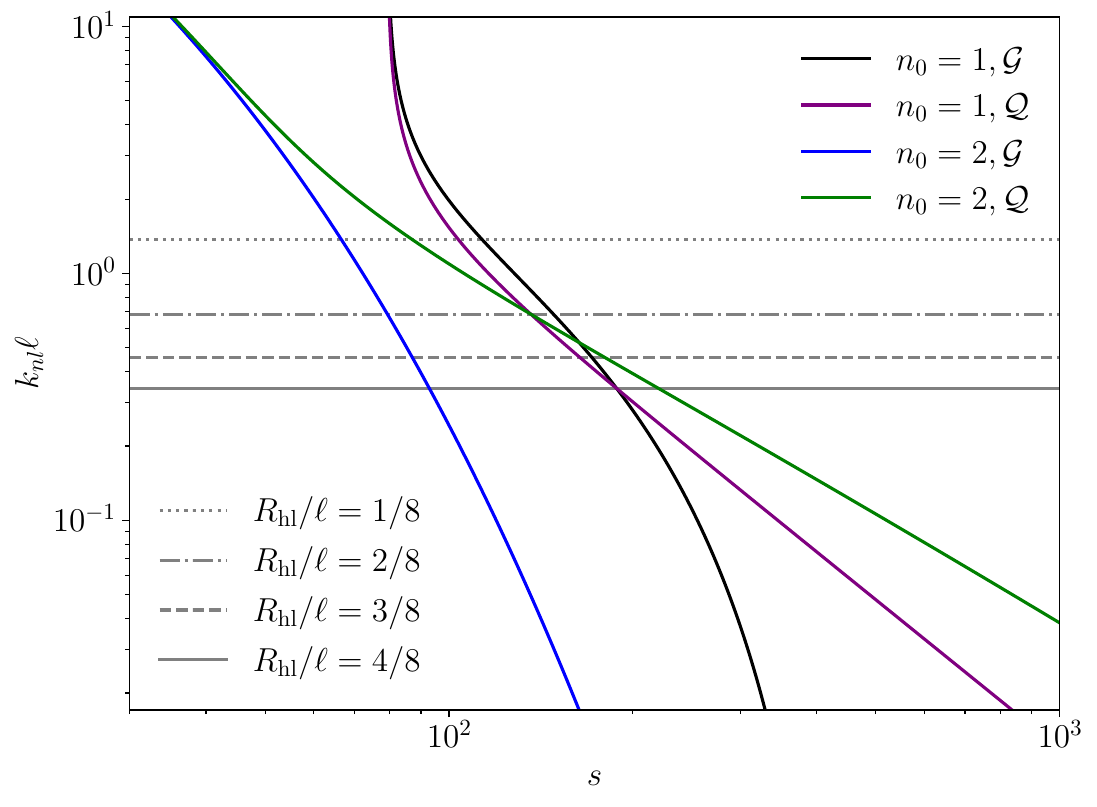}
        \caption{Evolution of the nonlinear wavenumber $k_{\rm nl}$ as a function of scalefactor.  Horizontal lines indicate show the characteristic halo sizes listed in Table~\ref{tab:hlradius}.  Steeper slopes correspond to more rapid merger histories and with more dilute halos.}
        \label{fig:knls}
    \end{figure}

    Our simulations imply that the amplitude of the correlation function depends sensitively on the evolution of the nonlinear wavenumber, which we show directly as a function of scalefactor in Fig.~\ref{fig:knls}.  For the simulations with matched initial conditions, we see that $k_{\rm nl}$ always changes more slowly in the $\mathcal{Q}$ simulation, consistent with halos being denser at all times.  For the matched final conditions, the evolution is more complicated as $k_{\rm nl}$ in the $\mathcal{Q}$ simulation is initially changing more rapidly than in the $\mathcal{G}$ simulation; although this soon reverses.  This is again consistent with the bottom panel of Fig.~\ref{fig:xiz_n1} where the correlation function ratio is initially reduced (albeit with the important caveat about numerical accuracy mentioned earlier) before subsequently becoming enhanced.

    Overall, we find that halos which evolve under the fractional $\mathcal{Q}$ potential are systematically denser than those evolved under the Newtonian force $\mathcal{G}$.  While it is possible to make ad-hoc changes to the initial conditions which partially reduce this, it is only possible at a single scalefactor.  By looking at larger $s$ in Fig.~\ref{fig:knls}, we expect that halos would become relatively denser in the $\mathcal{Q}$ simulation if we ran them for longer times.  Furthermore, we have shown that this increase in density correlates strongly with the rate of change of $k_{\rm nl}$ and so we expect it to be qualitatively similar for other partially screened forces.

\section{Nonlinear Screening}
    \label{sec:nonlinear}
    So far our results have all assumed that we can treat the scalar field linearly, which is only perturbatively true for the quartic (or higher power) potential.  In fact, we can already see that this may be an issue for the quartic self-interaction by inspecting the oscillation-averaged potential on large scales, $\mathcal{Q}_{k\bxi}\propto1/(k\ell)^{3/2}$ which diverges.  An even stricter condition that the oscillating potential remains small may be required, in which case $\delta_\psi$ needs to decay at least as fast as $\delta_\psi\propto(k\ell)^2$.  In our simulations this is not the case, and so we need to assume that the fermion perturbations decrease substantially faster on unresolved large scales, or that the scalar is otherwise regulated by relativistic effects or a small mass term, $\sim(1/2)m_\phi^2\phi^2$, in the potential.  
    
    While we can avoid such large scale issues with suitable assumptions, a more pressing concern is that the scalar field may become dynamically nonlinear during the simulation on resolved scales even when $\omega_\varphi^{-1}/\omega_\psi^{-1}\rightarrow0$.  Based on Eq.~\ref{eq:PoissonPhi1}, we can expect that this may occur when $\delta_\psi\sim(k\ell)\sim1$ as it is not suppressed by $(k\ell)^2$ nor $M_\phi^2$ when it oscillates to zero.  Below we argue that the typical halo size after structure formation is $R_{\rm hl}\lesssim\ell$ even under different assumptions about the internal dynamics of the halos.

    \subsection{Oscillating Fluctuations}
\begin{figure}
            \includegraphics[width=0.475\textwidth]{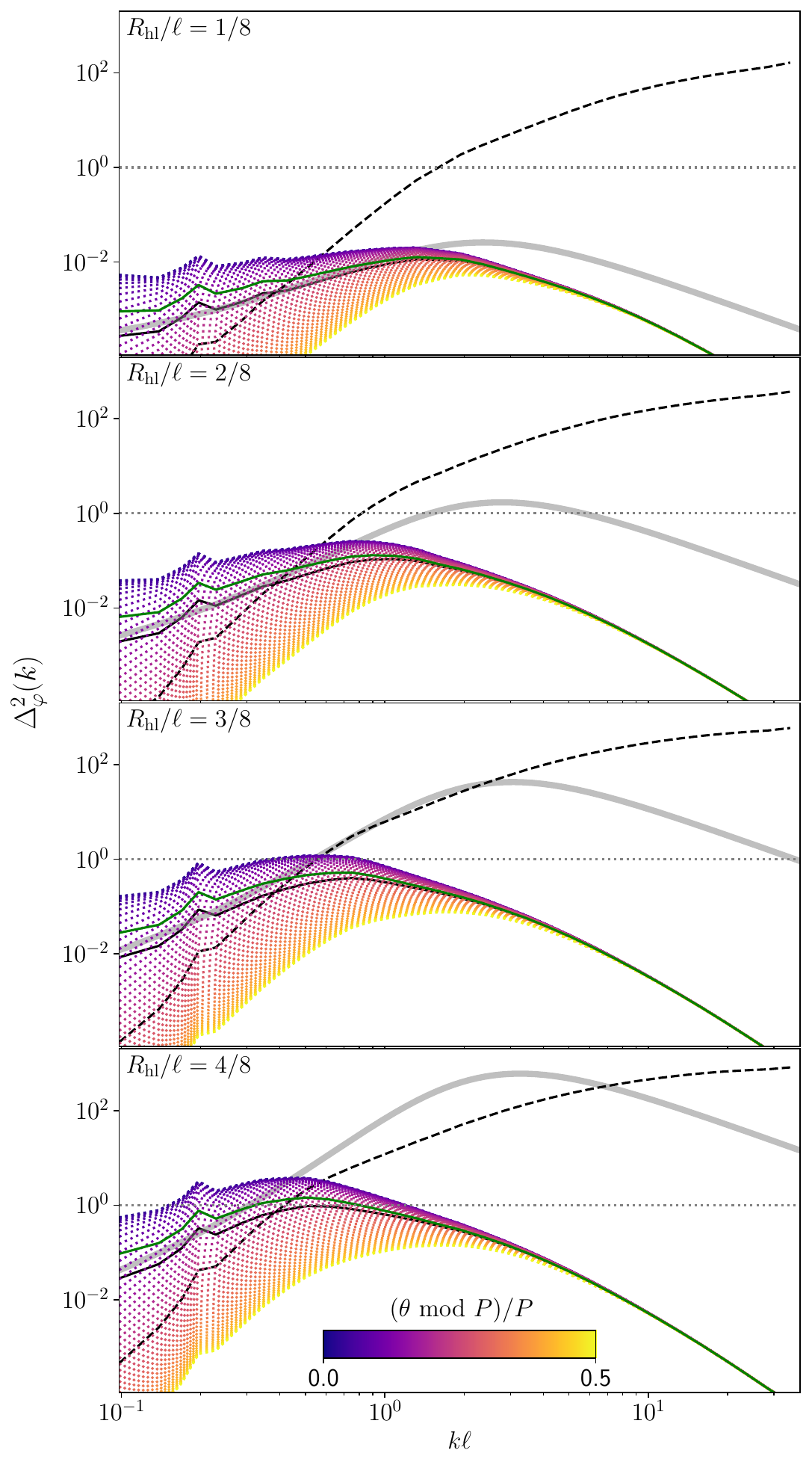}
            \caption{Power spectrum of the dimensionless scalar field perturbation as a function of oscillation phase for the quartic interaction and $n_0=2$ initial conditions.  The black curve shows the power spectrum of the oscillation averaged potential, while the grey curve shows its linear prediction.  The green curve is the oscillation average of the scalar power spectrum, which is quadratic in $\varphi$ and so differs from the black curve.  Scalar fluctuations around the halo become order unity when $R_{\rm hl}\lesssim\ell/2$.  For comparison, the fermion power spectrum is shown in the black dashed curve.}
            \label{fig:osc_pk}
        \end{figure}

        We now use our simulations to investigate when the quasilinear analysis breaks down, assuming that other dissipative or otherwise nonlinear processes are inefficient throughout.  The non-dimensional oscillating scalar perturbation is given in Fourier space as
        \begin{align}
            \frac{\varphi_1}{\bar\varphi}=\nu_1 &= -\frac{1}{3} \frac{\delta_\psi}{(k\ell)^2+\nu_0^2}.
        \end{align}
        from which we see that the background oscillations generate reciprocal oscillations in the perturbation - $\nu_1$ is largest when $\nu_0\sim0$ and smallest when $\nu_0^2$ is largest.  Of course, the simulations are designed not to resolve the oscillating potential; however, we can still estimate it by assuming that $\delta_\psi$ is constant over a period.  Under this approximation, we compute the dimensionless $\nu_1$ power spectrum $\Delta^2_\varphi$ as a function of the period.   We focus on the $n_0=2$ case as it is more consistent with a potential that is linear on large scales.  Fig.~\ref{fig:osc_pk} shows that the reciprocal oscillations are substantial and that the scalar power spectra becomes order unity when the characteristic halo size is $R_{\rm hl}\sim \ell/2$.

\begin{figure}
            \includegraphics[width=0.475\textwidth]{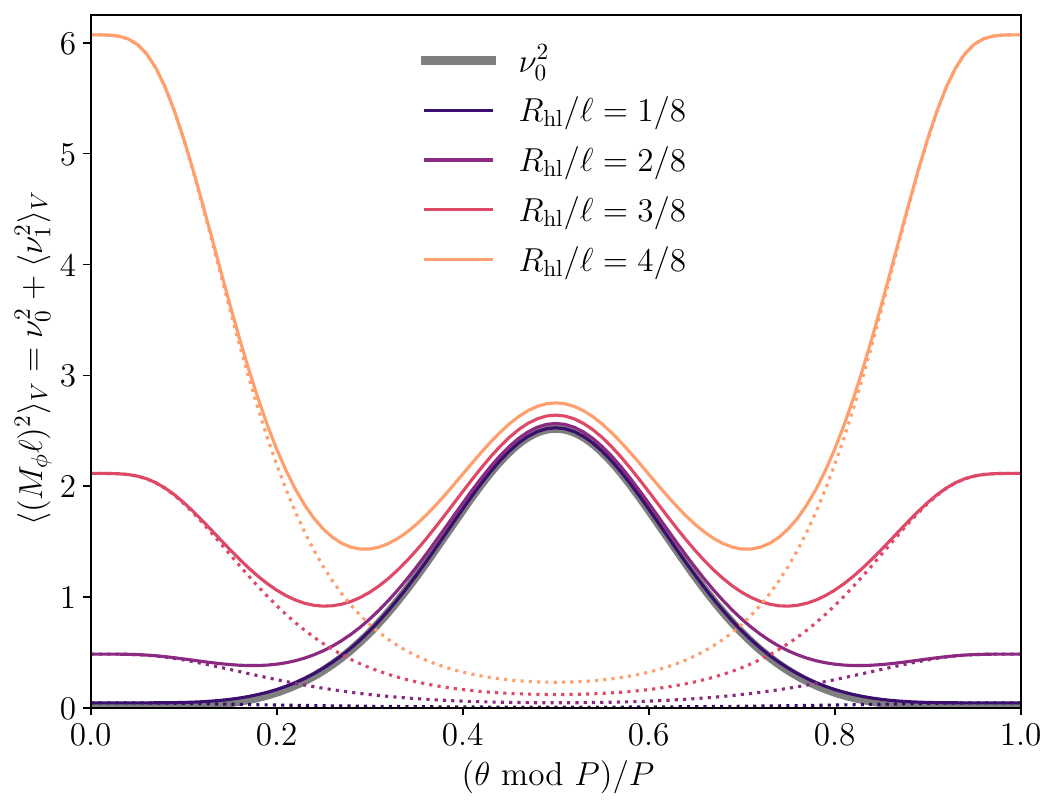}
            \caption{Scalar field mass for the quartic interaction as a function of background oscillation phase and typical halo size, which is monotonic with scalefactor, computed from the simulation with $\mathcal{Q}$ potential and $n_0=2$.  When the homogeneous component oscillates to zero the perturbation contribution is maximal whereas the reverse is true when the homogeneous contribution is large.  Dashed curves show just $\langle \nu_1^2\rangle_V$ and large values relative to $\nu_0^2$ indicate a breakdown in validity of the simulations.}
            \label{fig:M_phi2}
        \end{figure}

        The power spectrum quantifies fluctuations on certain scales and not necessarily typical values, i.e.,~the power being small for $k\ell\ll1$ implies a relatively constant but not necessarily small $\nu_1$.  
        To understand typical amplitudes, we compute the scalar field mass, $\left(M_\phi\ell\right)^2 = \nu_0^2 + \nu_1^2$, and show its RMS value, $\langle(M_\phi\ell\rangle^2_V$ as a function of the oscillation phase in Fig.~\ref{fig:M_phi2}.  We find that for $R_{\rm hl}/\ell\gtrsim 1/8$ the scalar field is no longer effectively massless for part of the oscillation, indicating the end of validity for the quasilinear simulations.
        
    \subsection{Equilibrium Behaviour}
        Without simulations evolving the nonlinear Klein-Gordon equation for the scalar field, it is difficult to move beyond the previous analysis quantitatively.  Nonetheless, we can consider more qualitative arguments about what will happen to the halos under conditions where they can equilibrate \citep{bib:Sasaki2024}.  Then, energy will be equipartitioned between the scalar binding energy and its self-interaction energy,
        \begin{align}
            \frac{y^2 N_{\rm hl}^2}{R_{\rm hl}} \sim V(\phi)R_{\rm hl}^3
        \end{align}
        where $N_{\rm hl}=M_{\rm hl}/m_\psi$ is the number of fermions in a halo of radius $R_{\rm hl}$ and we neglect factors of order unity.  The expression for the binding energy is based on analogy with the gravitational case, and so is only valid provided the range of the force, $M_\phi^{-1}$ is larger than the halo.  For the quartic potential, the scalar mass is $M_\phi^2\sim\lambda\phi^2$ yielding
        \begin{align}
             \left(\frac{N_{\rm hl}}{n_i a_i^3\ell^3}\right)^2\sim M_\phi^4 R_{\rm hl}^4
        \end{align}
        where we also used that $\sqrt{\lambda} y\sim 1/(n_ia_i^3\ell^3)$.
        As the halo continues to grow, it can reach $N_{\rm hl}\sim n_ia_i^3\ell^3$,  implying $M_\phi\sim1/R_{\rm hl}$ and that the fifth force is screened at the size of the bound object.
        
        This behaviour is essentially the chameleon mechanism in modified gravity scenarios, where a similar interaction with $V_{\rm eff}(\phi)$ has already been used to implement density-dependent screening \citep{bib:Gubser2004}.  Compact systems of this type in the early Universe were studied in Ref.~\citep{bib:Lu2025}, who found that the degeneracy pressure supported Fermi Balls can collapse into primordial black holes once they become sufficiently large.  Since our results suggest that growth during the prior dilute phase may be less efficient in producing $N_{\rm hl}>1/\sqrt{\lambda}y$, another merger mechanism may be necessary to form objects of sufficient size to collapse.  
    
\section{Conclusion}
    We have run high-resolution N-body simulations of early Universe fermion clustering under different types of Yukawa forces.  We considered two regimes - one where we assume that the dynamics are on scales much smaller than a generalized Compton wavelength, relevant for any scalar field potential, and one where the dynamics are on scales comparable to the screening length.  For the latter, we solved a fractional equation valid when the scalar field has a quartic potential and is oscillating on fast timescales compared to the fermions.  
    
    We have two main conclusions from the high-resolution simulations.  The first is that, when compared at similar mass scales, halos that form under a screened force are significantly denser.  We identified that this is closely associated with the rate of change of the nonlinear scale - that is, how fast halos are growing by mergers.  We therefore expect our conclusion to be qualitatively correct for other screened potentials too.  Our second conclusion, specific to the quartic self-interaction, is that reciprocal oscillations in the perturbed scalar field become important when the halo radius becomes comparable to the background screening length. If the scalar field mass becomes large, it may limit the size of halos to around the associated Compton length, although showing this directly would require simulations which evolve nonlinear equations for the scalar.  However, even if the scalar force becomes completely screened, the initial instability still leads to correlations on scales much larger than $\ell^{-1}$ which could then be constrained by observations similarly to the quadratic case \citep{bib:Costa2025,bib:Graham2025}.  Further incorporating radiative and nonlinear processes is clearly an important target for future analyses.
    
\section{Acknowledgements}
I am grateful to Misao Sasaki for substantial discussions and suggestions related to the analysis and interpretation of the simulations, as well as for many comments on the paper itself.  I furthermore acknowledge valuable discussions with Zachary Picker on Fermi balls, and with Ravi Sheth about the shell theorem and radiation era initial conditions.  This work has made use of Matplotlib \citep{bib:Hunter2007}, SciPy \citep{bib:Virtanen2020}, and NumPy \citep{bib:Harris2020} as well as the Astrophysics Data System which is funded by NASA under Cooperative Agreement 80NSSC21M00561.

\bibliography{msbib}

\appendix
\section{Expansion in $\omega_\varphi^{-1}/\omega_\psi^{-1}$}
    \label{app:expansion}
    
\subsection{Multiscale Expansion in $\varphi$}    
    \label{ssec:multiscale}
      We define $\varphi=\bar\varphi\nu$ and substitute into \ref{eq:varscalareqn} to obtain
        \begin{align}
            \frac{\bar\varphi}{y n_i a_i^2} \left[ \nu''-\nabla^2\nu \right] + \nu^{3} + 1 + \delta_\psi = 0.
        \end{align}
        It is immediately that there is no free parameter to keep the scalar field linear, and furthermore, while we expect the Laplacian term $-\nabla^2\nu$ to smooth the scalar field on small scales, it can still be the case that $|\nu|^3\sim\delta_\psi\gg|\nu|\gtrsim1$.  
        
        Rather than expanding to higher orders $\varphi=\varphi_0+\varphi_1+\varphi_2+...$, let us instead make use of the freedom to choose our time and space coordinates to make a non-perturbative expansion in $\omega_\varphi^{-1}/\omega_\psi^{-1}$.  Based on the quasilinear dynamics, we expect that the fermion overdensity grows on scales characterized by the scalar field oscillation frequency $\vec{\chi}=\omega_\varphi \vec{x}\sim\vec{x}/\ell$, but on timescales set by the fermion evolution $s=\omega_\psi\eta$, which we specify explicitly as $\delta_\psi=\delta_+(s,\vec{\chi})$.  It is natural then to look for a scalar term on the same scales, i.e., $\nu\sim\nu_+(s,\vec{\chi})$.  However, this does not match the homogeneous component which instead evolves on fast timescales $\theta=\omega_\varphi\eta$ and long wavelengths $\vec{\Lambda}=\omega_\psi\vec{x}$, $\nu\sim\nu_\times(\theta,\vec{\Lambda})$.  We therefore consider $\nu(\eta,\vec{x})=\nu_\times(\theta,\vec{\Lambda}) + \nu_+(s,\vec{\chi})$, separating the equations into
        \begin{align}
            \frac{\partial^2}{\partial\theta^2}\nu_\times + \nu_\times^3 &+ 1 - \left(\frac{\omega_\varphi^{-1}}{\omega_\psi^{-1}}\right)^2\nabla^2_\Lambda\nu_\times \nonumber \\ 
            & +3\nu_\times \langle \nu_+^2\rangle_V + \langle \nu_+^3 \rangle_V = 0. \label{eq:nu_times}
        \end{align}
        and
        \begin{align}
            - \nabla^2_\chi \nu_+ &+ 3 \nu_\times^2\nu_++ \delta_+ + \left(\frac{\omega_\varphi^{-1}}{\omega_\psi^{-1}}\right)^2 \frac{\partial^2}{\partial s^2} \nu_+ \nonumber\\ &+ 3 \nu_\times (\nu_+^2-\langle\nu_+^2\rangle_V) + (\nu_+^3-\langle\nu_+^3\rangle_V)  = 0, \label{eq:nu_plus}
        \end{align}
         where we have placed terms with the expectation that $\varphi_0\approx\bar\varphi\nu_\times$ represents the homogeneous background and $\varphi_1\approx\bar\varphi\nu_+$ is the fluctuation.  

        However, we see directly that the nonlinear $\nu^3$ term necessarily mixes different time and length scales, even when $\omega_\varphi^{-1}/\omega_\psi^{-1}\rightarrow0$.  If we assume $\langle \nu_+^2 \rangle_V\ll1$ then $\nu_+$ becomes the oscillating background $\nu_0$ while Eq.~\ref{eq:nu_plus} retains a linear mixed term $3\nu_\times^2\nu_+$ which can be further averaged to obtain the fractional force law in Eq.~\ref{eq:Qpoisson}.  On the other hand, if $\nu_+^3\gg|\nu_\times|\sim\langle \nu_+^2 \rangle_V\gg1$ as we might expect in a deeply nonlinear regime with $\delta_+\gg1$, then Eq.~\ref{eq:nu_plus} decouples from $\nu_\times$,
        \begin{align}
        - \nabla^2_\chi \nu_+ + \nu_+^3 + \delta_+ \approx 0.
        \end{align}
        It is only the first regime that can be modelled by the simulations.

\subsection{Friction and Tidal Effects in $\psi$}
    \label{app:corrections}
    
    In this section we compute the lowest order correction terms to the oscillation-averaged dynamics.  However, we continue to ignore relativistic effects (i.e.,~$\varphi_1''$), nonlinear in $\varphi$ terms such as $\varphi_0\varphi_1^2$ or $\varphi_1^3$ (and others that we shall soon encounter), as well as gravitational effects (e.g.,~the horizon scale or the onset of the matter era).

    We first compute the correction to Eq.~\ref{eq:osc_avg_x} which is related to the oscillating fermion mass \citep{bib:Domenech2023},
    \begin{align}
        m_{\rm eff}=m_\psi + y \frac{\varphi}{s/s_i}\approx m_\psi\left(1+y \frac{\bar\varphi}{m_\psi} \frac{\nu_0}{s/s_i}\right)      
    \end{align}
    where we neglected the term proportional to $\varphi_1$.
    We further assume that $y\bar\varphi/m_\psi\ll1$ such that we can treat this as always perturbative and decaying.  Then we have that
    \begin{align}
        \frac{\vec{x}(\eta+\Delta\eta)-\vec{x}(\eta)}{\Delta\eta} &= \frac{1}{\Delta\eta}\int_\eta^{\eta+\Delta\eta} \frac{\vec{p}}{a m_{\rm eff}}d\eta' \\
        &\approx \frac{\vec{p}}{m_\psi}\frac{1}{\Delta\eta}\int_\eta^{\eta+\Delta\eta} \frac{1}{a}\frac{d\eta'}{1+y \frac{\bar\varphi}{m_\psi} \frac{\nu_0}{s/s_i}} \\
        &\approx \frac{\vec{p}}{a m_\psi} \left(1-y \frac{\bar\varphi}{m_\psi} \frac{\langle\nu_0\rangle}{s/s_i}\right) \\
        &\equiv \frac{\vec{p}}{a m_\psi} \mathcal{M} \label{eq:dx_m}
    \end{align}
    where $\langle\nu_0\rangle$ is an order 1 number.  The correction $\mathcal{M}-1$, assumed to start small, decays even further from relevance. 

    We next compute the corrective terms to Eq.~\ref{eq:osc_avg_v}.  In this case we have that 
    \begin{align}
        \frac{\vec{p}(\eta+\Delta\eta)-\vec{p}(\eta)}{\Delta\eta} = -\frac{a_i y}{\Delta\eta}\int_\eta^{\eta+\Delta\eta}\vec{\nabla}\varphi_1(\eta',x(\eta')) d\eta'
    \end{align}
    and we should evaluate trajectories based on Eq.~\ref{eq:dx_m}, $\vec{x}(\eta'\le \eta+\Delta\eta)\approx \vec{x}(\eta) + \vec{p}/(a m_\psi)(\eta'-\eta) \mathcal{M}$, corresponding to the Born approximation.  Then we have that
    \begin{align}
        \varphi_1(\eta',\vec{x}(\eta')) &\approx \varphi_1(\eta',\vec{x}(\eta))\nonumber\\ &+ \frac{\vec{p}(\eta)}{a(\eta) m_\psi}\mathcal{M}(\eta) (\eta'-\eta)\vec{\nabla}\varphi_1(\eta',\vec{x}(\eta))
    \end{align}
    and so
    \begin{widetext}
    \begin{align}
        \frac{\vec{p}(\eta+\Delta\eta)-\vec{p}(\eta)}{\Delta\eta} = -\frac{a_i y}{\Delta\eta}\int_\eta^{\eta+\Delta\eta} \left( \vec{\nabla}\varphi_1 + \frac{\mathcal{M}(\eta)}{a(\eta) m_\psi}(\vec{p}(\eta)\cdot\vec{\nabla}) \vec{\nabla}\varphi_1 (\eta'-\eta) \right)d\eta'.
    \end{align}
    \end{widetext}
    The first term under the integral is the oscillation averaged force already found, while the second represents a tidal correction.  This correction depends explicitly on $\eta$, e.g.,~the scalar field value,
    \begin{align}
        \langle \delta\mathcal{Y}_{k\xi} \rangle &\propto \frac{1}{\Delta\eta}\int_\eta^{\eta+\Delta\eta} \frac{\eta'-\eta}{k^2+\xi^{-2}(\eta')} d\eta' \\
        &=\frac{1}{\Delta\eta}\int_0^{\Delta\eta} \frac{y}{k^2+\xi^{-2}(\eta+y)} dy
    \end{align}
    with $y=\eta'-\eta$.  One possibility is to perform another oscillation average (e.g., because $d\eta\gg\Delta\eta$).  If we do this we get
    \begin{widetext}
    \begin{align}
        \langle\langle \delta\mathcal{Y}_{k\xi} \rangle\rangle =\frac{1}{\Delta\eta^2}\int_0^{\Delta\eta}dx\int_0^{\Delta\eta}dy\frac{y}{k^2+\xi^{-2}(\eta+x+y)} \frac{1}{(\eta+x)a_i/\eta_i}\mathcal{M}(\eta+x)\approx \frac{1}{2}\frac{\Delta\eta}{a}\mathcal{M}\mathcal{Q}_{k\bxi}
    \end{align}
    \end{widetext}
    where we assumed $\eta\gg\Delta\eta\ge x$.  
    
    The lowest order equations of motion therefore are,
    \begin{align}
        \frac{d\vec{x}}{d\tau} = \frac{\vec{p}}{m_\psi}\mathcal{M}
    \end{align}
    and 
    \begin{align}
        \frac{1}{m_\psi}\frac{d\vec{p}}{d\tau} = -\vec{\nabla}\Phi + \vec{a}_{\rm T}
    \end{align}
    where the tidal acceleration is given by
    \begin{align}
        \vec{a}_T&=-\frac{1}{2}\frac{\Delta\eta}{a}  \mathcal{M}\left(\vec{v}\cdot\vec{\nabla}\right)\vec{\nabla}\Phi \\ 
        &= -\frac{1}{2} \left( P (a_i^2H_i)^{-1} \mathcal{M}\frac{\omega_\varphi^{-1}}{\omega_\psi^{-1}}\right)\left(\frac{\vec{v}}{s}\cdot\vec{\nabla}\right) \vec{\nabla}\Phi.
    \end{align}
    This tidal term does not decay (initially at least, as $\vec{v}\propto s$), unlike the friction $\mathcal{M}\rightarrow1$.  As discussed in \S~\ref{ssec:multiscale}, the oscillation-averaged force equation is already correct to order $(\omega_\varphi^{-1}/\omega_\psi^{-1})^2$ assuming quasilinearity.  Therefore, in the limit where $y\bar\varphi/m_\psi$ and $\omega_\varphi^{-1}/\omega_\psi^{-1}$ are negligible we recover the equations used for the simulations.

\section{Primordial Displacements}
    \label{app:primdisp}

    The primordial displacement field $\vec{\Psi}_0$ introduces a conceptual ambiguity for particle initialization.  We have considered two possible particle initial conditions to deal with this.  In the first, we neglect the ambiguity completely and assign particle displacements analogously to traditional large scale structure simulations:
    \begin{align}
        \vec{\Psi}_p =\vec{\Psi} \approx -i \frac{\vec{k}}{k^2}\delta_\psi(s_0)\left[1+\frac{1}{4}(-k^2\mathcal{Q}_{k\bxi})(s-s_0)\right]. \label{eq:displacement1}
    \end{align}
    In the second possibility, we encode only the relative displacement between $s_i$ and $s_0\rightarrow0$,
    \begin{align}
        \vec{\Psi}_p = \vec{\Psi}-\vec{\Psi}_0 \approx i \frac{\vec{k}}{k^2}\delta_\psi(s_0)\left[\frac{1}{4}(-k^2\mathcal{Q}_{k\bxi})(s-s_0)\right]. \label{eq:displacement2}
    \end{align}
    With this second choice however, we still need to encode the primordial fluctuations at $s_0$ which we can do by adding an additional fluctuation to the particle mass, $\delta m_p/m_p$.  In either case the velocities are set by 
    \begin{align}
    s\frac{d}{ds}\vec{\Psi}_p\approx -i\vec{k}k^{-2}\delta_\psi(s_0)(-k^2\mathcal{Q}_{k\xi})\frac{s}{4},
    \end{align}
    although only for the second choice are they proportional to the displacement field for $s_0=0$.  We note that this situation is somewhat analogous to one encountered in primordial black hole simulations; however, in that case the black holes are initially Poisson distribution and so particles can be placed randomly rather than on a lattice \citep{bib:Inman2019}.
    
\begin{figure}
        \includegraphics[width=0.45\textwidth]{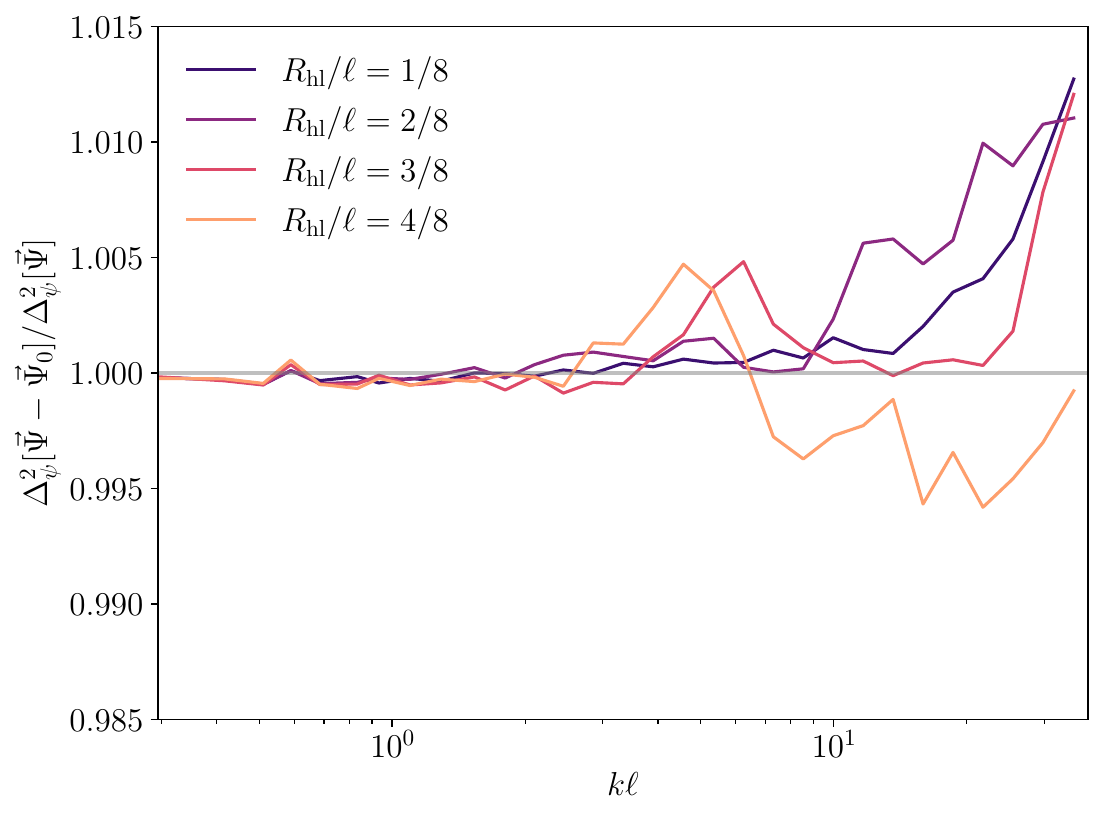}
        \caption{Ratio of power spectrum computed with mass fluctuations and displacements to that where there are only displacements.  The difference are generally sub-percent, although it is difficult to extrapolate to even smaller scales.}
        \label{fig:pkic}
    \end{figure}

    Using individual particle mass fluctuations has only seen limited use in cosmological simulations (e.g.,~\citep{bib:Hahn2021}), although having a pair of masses is common for multi-species simulations (e.g.,~\citep{bib:Liu2023}).  The simplest way to encode the density field is to have one particle at the center of each grid cell, and assign that particle's mass fluctuation to be $\delta m_p/m_p=\delta_\psi(s_0)$.  However, since this requires at least four times more particles than our preferred lattice conditions, our default simulation setup was not possible.  Instead we reduced the grid size to $n_c=256^3$ cells with the same number of particles (i.e.,~just over seven times fewer particles in total than the default setup).  We ran two simulations, one with only displacements given by Eq.~\ref{eq:displacement1} and one with both mass fluctuations and displacements given by Eq.~\ref{eq:displacement2}, showing their ratio in Fig.~\ref{fig:pkic}.  We find that on small scales there is around a $\sim1\%$ difference between the two simulations.  Unfortunately, it is somewhat difficult to interpret how this impacts our results due to the changing particle number, PM grid size, and ratio of $n_\ell$ to $n_c$.  However, our results are principally comparisons between simulations which should be similarly affected by this ambiguity and so we do not expect our interpretations to change.  Determining how to create more accurate initial conditions in the radiation era is certainly worthy of further study.

\section{Convergence Tests for $\Xi$}
    \label{app:convergence}

\subsection{Particle Interpolation}
    \label{appsub:xipartinterp}

\begin{figure}
        \includegraphics[width=0.45\textwidth]{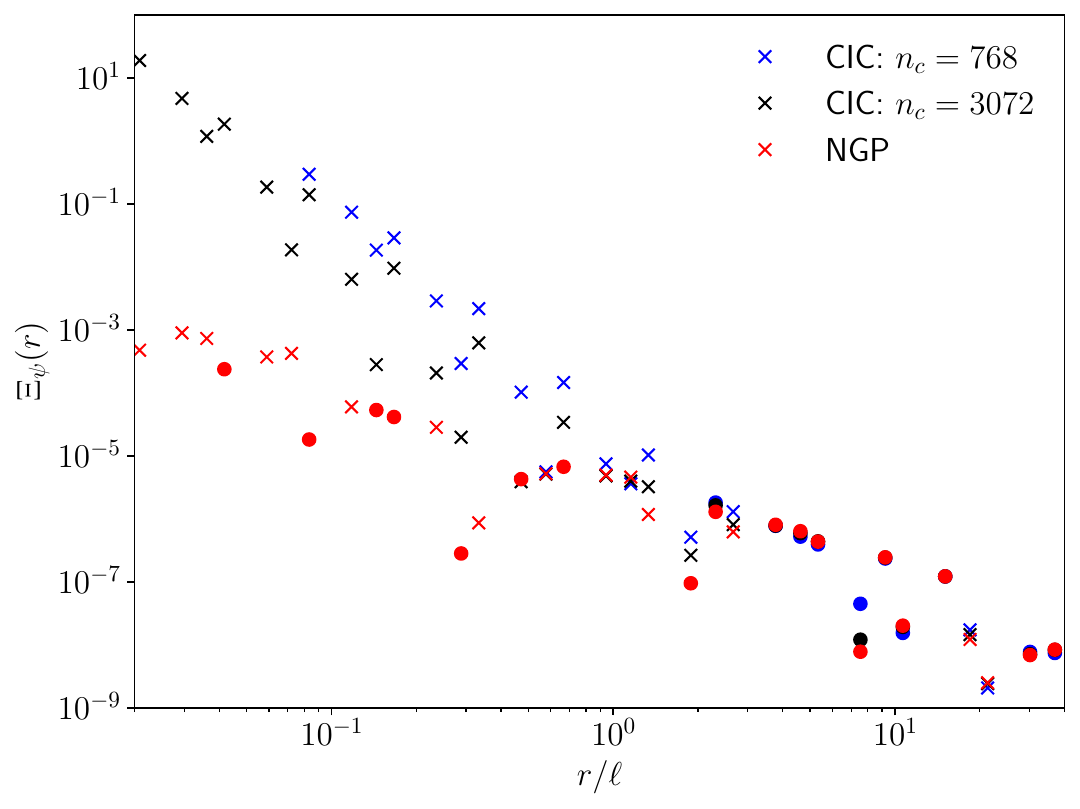}
        \caption{Correlation function of particles with random coordinates using different interpolation schemes.  When interpolation is done using the NGP method the results are independent of grid size.}
        \label{fig:xi_pn}
    \end{figure}

    In this appendix we discuss how the correlation function calculation depends on the particle interpolation method.  As a demonstration, we compute the correlation function of particles with random coordinates using either the CIC or NGP interpolation method, showing the result in Fig.~\ref{fig:xi_pn}.  While a true Poisson distribution is expected to have zero correlation, we find there is a numerical correlation which is substantially larger when using CIC interpolation.  
    
\begin{figure}
        \includegraphics[width=0.45\textwidth]{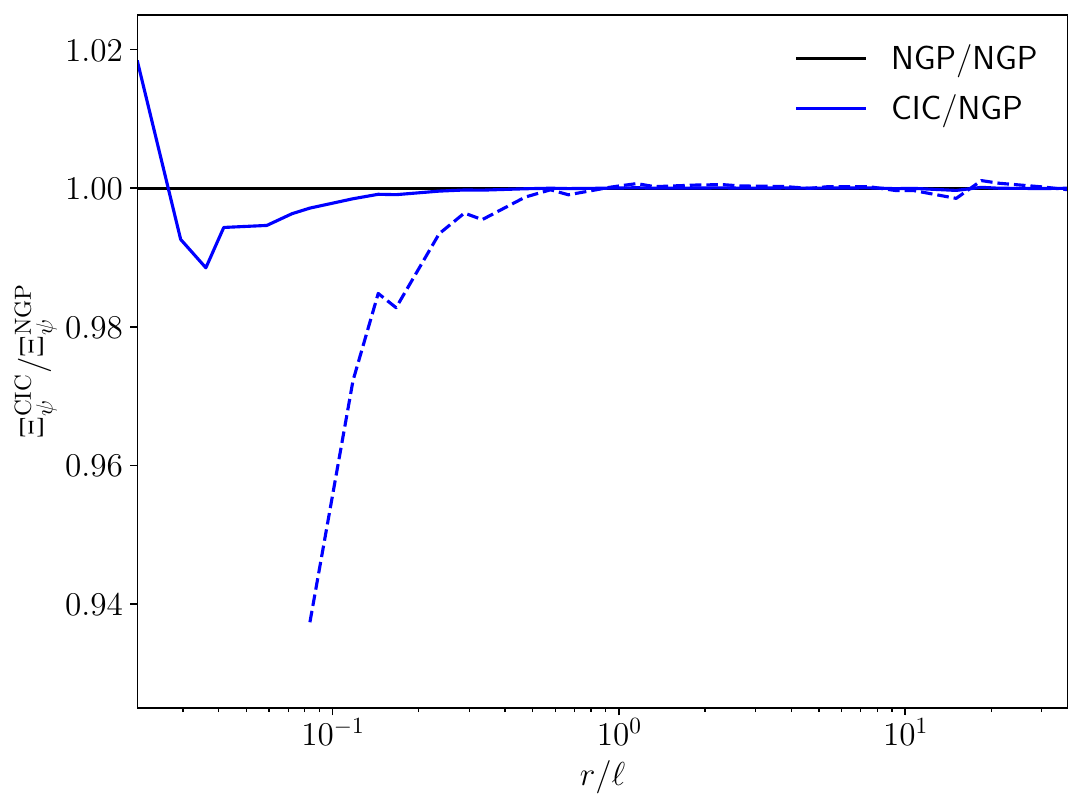}
        \caption{Ratio of correlation functions computed using CIC to NGP at the final redshift of the $n_0=2$ $\mathcal{G}$ simulation.  The solid purple line is computed with a grid size of $n_c=3072$ cells per dimension while the  dashed purple line uses $n_c=768$.  By construction an NGP calculation with $n_c=768$ is equivalent to $n_c=3072$ for grid cells that overlap.}
        \label{fig:xi_ngp}
    \end{figure}

    In both cases however, the numerical result is substantially less than the physical correlations we find in the simulation.  At the final redshift of the simulations, we find that it makes very little difference whether we use NGP or CIC provided the number of grid cells is large enough.  We show this in Fig.~\ref{fig:xi_ngp} where the differences between CIC and NGP are quite small for the $n_c=3072$ grid.  However, the differences are not small when a $n_c=768$ grid is used suggesting that NGP interpolation provides a more robust estimate.

\subsection{Self-Similarity}
    \label{appsub:sftest}

    While we can compare our simulations with linear theory predictions on linear scales, it is much more difficult to test the accuracy of the simulation and analysis methods on nonlinear scales, especially as we have no other implementations to compare against.  An alternative way to determine accuracy is to exploit self-similarity \citep{bib:Joyce2021}.  This approach makes use of the fact that the only physical scale in an initially scale-free simulation is the nonlinear scale.  Provided results are appropriately re-scaled by a quantity that varies with the nonlinear scale, we can establish the degree to which the simulations are converged.

\begin{figure}
        \includegraphics[width=0.45\textwidth]{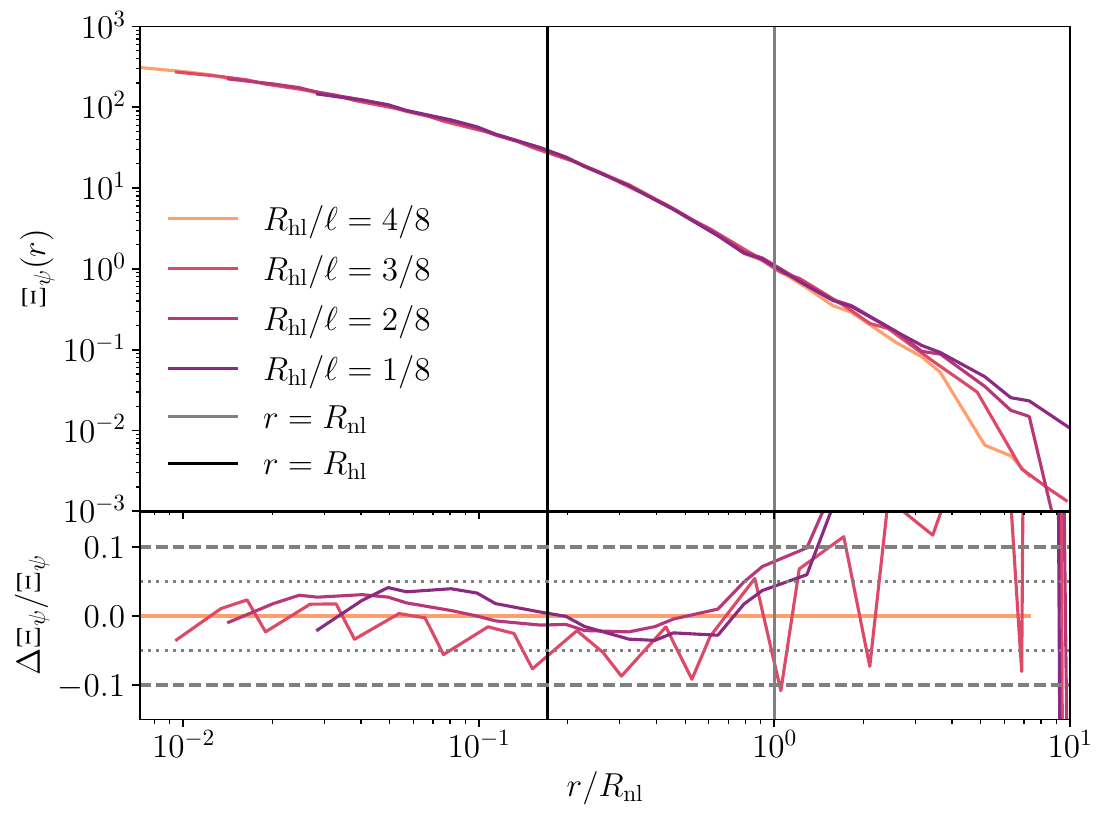}
        \caption{Correlation function of scale-free $\mathcal{G}$ with $n_0=2$ simulation, which has been re-scaled to show self-similarity.}
        \label{fig:xisf_ngp}
    \end{figure}

    Our simulation with $n_0=2$ and $\mathcal{G}$ represents an example of a simulation that should be scale-free.  We show the re-scaled correlation function in Fig.~\ref{fig:xisf_ngp}, taking $R_{\rm nl}=R_{\rm hl}\Delta_{\rm cr}^{1/3}=k_{\rm nl}^{-1}$ as an appropriate re-scaling variable.  In general, we find that our results appear to be converged to around 5\% up to the halo radius and to better than 10\% up to the nonlinear scale.  Afterwards, they differ more substantially from self-similarity, likely due to the effects of the finite box size.  

    Another very noticeable effect is the saw-like pattern seen in the bottom panel for $R_{\rm hl}/\ell=3/8$.  This is not due to convergence of our simulations, but rather a systematic effect in the calculation of $\Xi$, specifically due to the three different shifts $\sqrt{1}$, $\sqrt{2}$ and $\sqrt{3}$ used in the calculation.  The resulting artifact cancels exactly when the radii of the $4/8$ correlation function is interpolated to the $1/8$ and $2/8$ cases as the points remain aligned, but does not for the $3/8$ case.  We therefore treat the size of the saw pattern as an estimate of the numerical error in our calculation, which appears to be less than 10\% out to $R_{\rm nl}$.

\subsection{Numerical Parameters}
    \label{appsub:numericalparams}

\begin{figure}
        \includegraphics[width=0.45\textwidth]{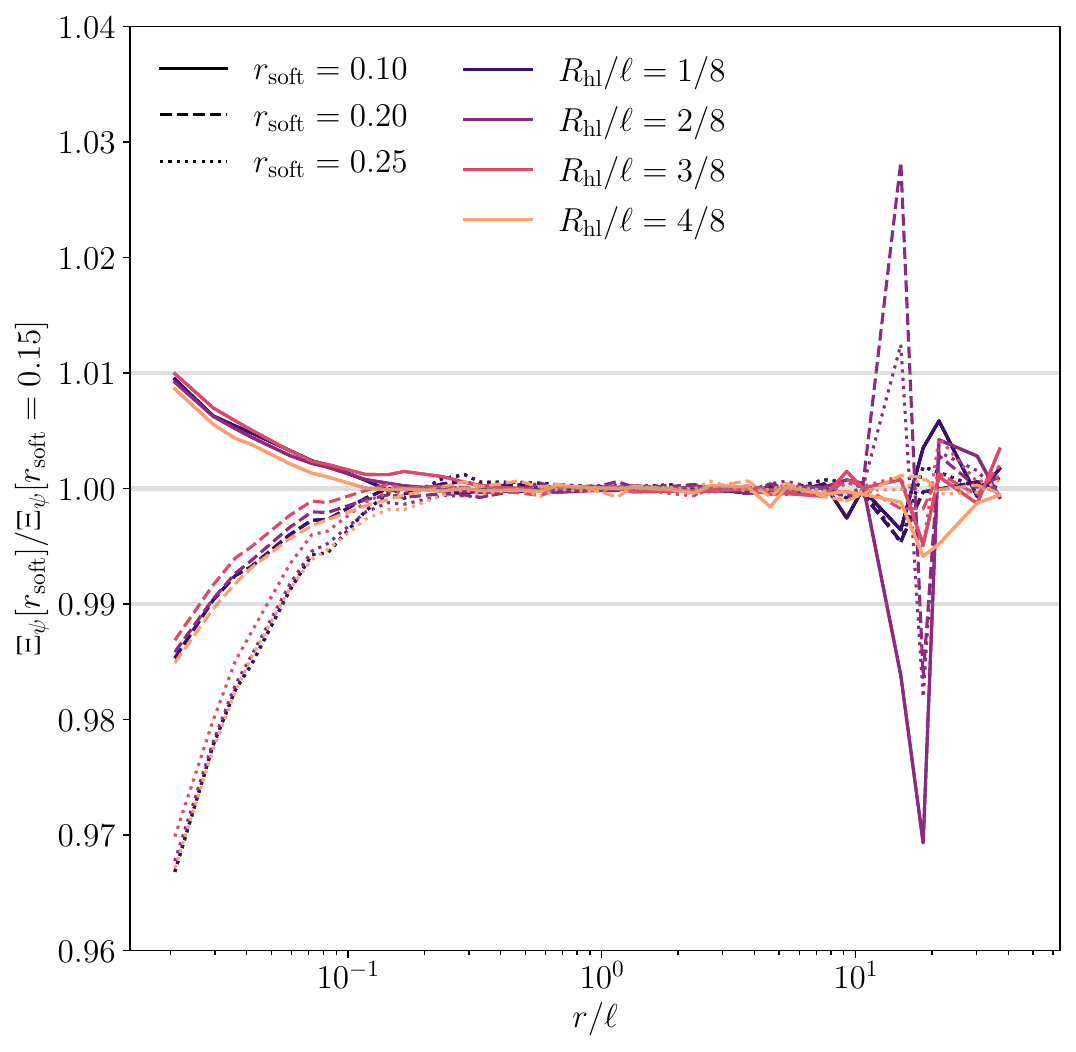}
        \caption{Variation of the correlation function for the $\mathcal{G}$ simulation with $n_0=2$ with respect to softening length.  The denominator of the ratio is shown directly in Fig.~\ref{fig:xiz_n2}.}
        \label{fig:xiz_soft}
    \end{figure}

    We lastly wish to test the effects of numerical parameters on our calculation.  We first consider the force softening length: we ran three further $\mathcal{G}$ simulations with $n_0=2$, but using $r_{\rm soft}=0.1,\ 0.2,$ and $0.25$.  Fig.~\ref{fig:xiz_soft} shows the ratio of the correlation function with varying softening length to the one in Fig.~\ref{fig:xiz_n2} .  We find that our results are generally converged to around $\sim1\%$ on all scales.  However, it does seem to be the case that in all cases deviations are beginning around $r\sim10^{-1}\ell$, which corresponds to the grid scale.  By comparing with Fig.~\ref{fig:facc} we see that the deviation starts occurring on scales comparable but smaller than where the largest force error is.  Thus, differences due to the PM and PP force matching remains a plausible explanation.  Since $\mathcal{Q}\approx\mathcal{G}$ on scales relevant to force softening, we do not expect it to directly change between the sets of simulations, unless in conjunction with other effects.

    \begin{figure}
        \includegraphics[width=0.45\textwidth]{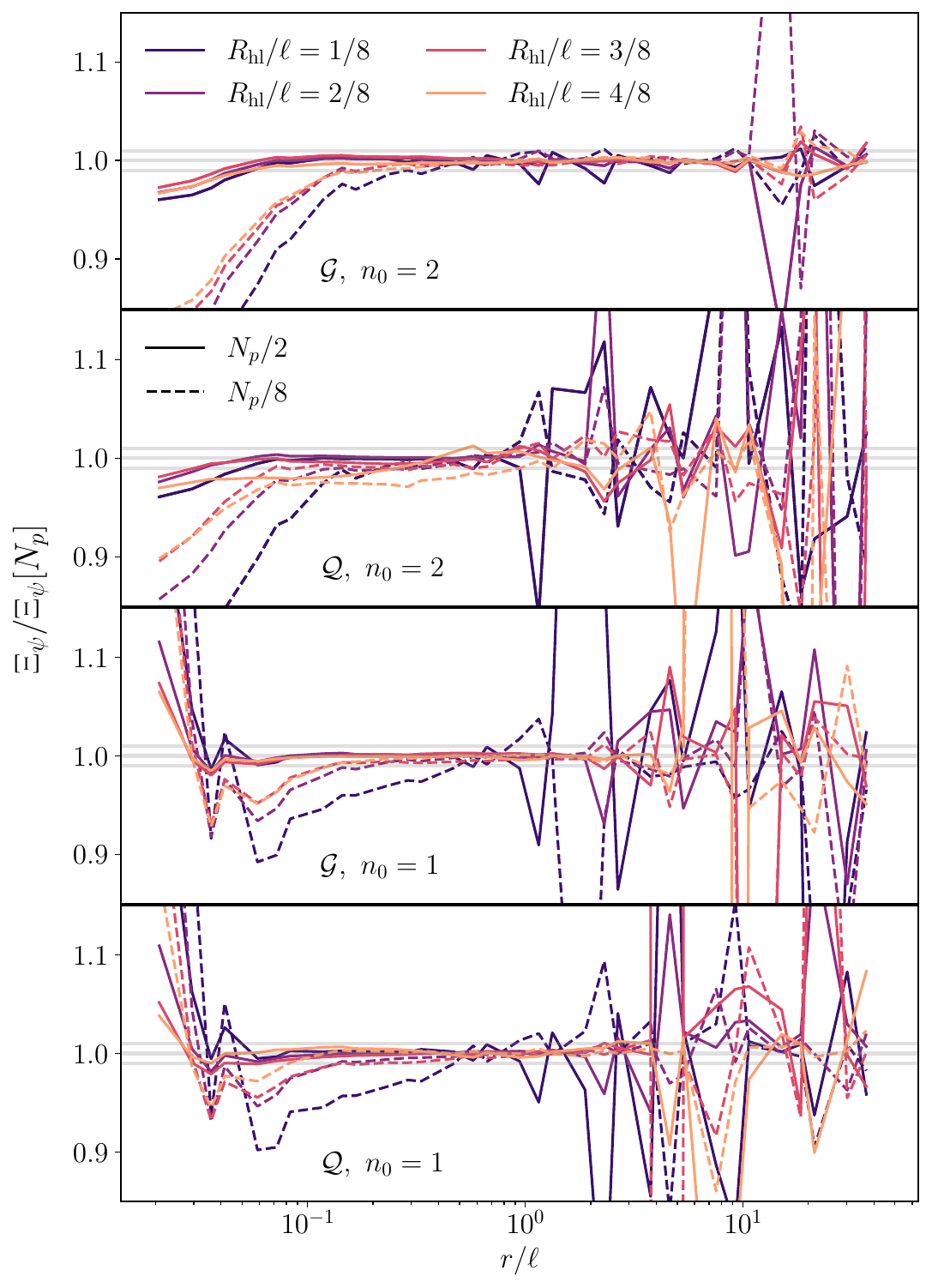}
        \caption{Variation of the correlation function with respect to particle number for each of the four cases we simulated.  We find general convergence to be $\sim1\%$ in most cases.  The large fluctuations on scales $r\gtrsim\ell$ appear to be due to the calculation method as they do not appear in the power spectrum.}
        \label{fig:xiz_np}
    \end{figure}  

    The second test we perform tests the convergence with respect to N-body particle number.  The simplest way to change particle number is to reduce the particle number by 2 per dimension, or a factor of 8 total.  However, this is a rather dramatic change, and so we have also run a set of simulations where we change the lattice configuration to a simple cubic lattice (see \citep{bib:Joyce2009} for a visualization), which corresponds to a factor of 2 reduction in particle number.  We show the ratio between the correlation function with reduced particle number to the ones shown in Figs.~\ref{fig:xiz_n2} and ~\ref{fig:xiz_n1}.  On scales $r\lesssim\ell$, we find that the simulations with $N_p/2$ are generally converged to better than 1\% to $r\sim10^{-1}\ell$.  One exception to this is the correlation function of the last redshift of the $\mathcal{Q}$ simulation with $n_0=2$.  Here we observe a larger and more scale-dependant change in the ratio.  The origin of this is not clear, although one plausible explanation is that this simulation runs for to a substantially higher scalefactor, and so numerical issues may become more pronounced.  For the $N_p/8$ ratio, we find worse convergence in all cases, reaching around $10\%$ at $r\sim10^{-1}\ell$, with earlier times show the greatest deviation in all cases.  

\begin{figure}
        \includegraphics[width=0.45\textwidth]{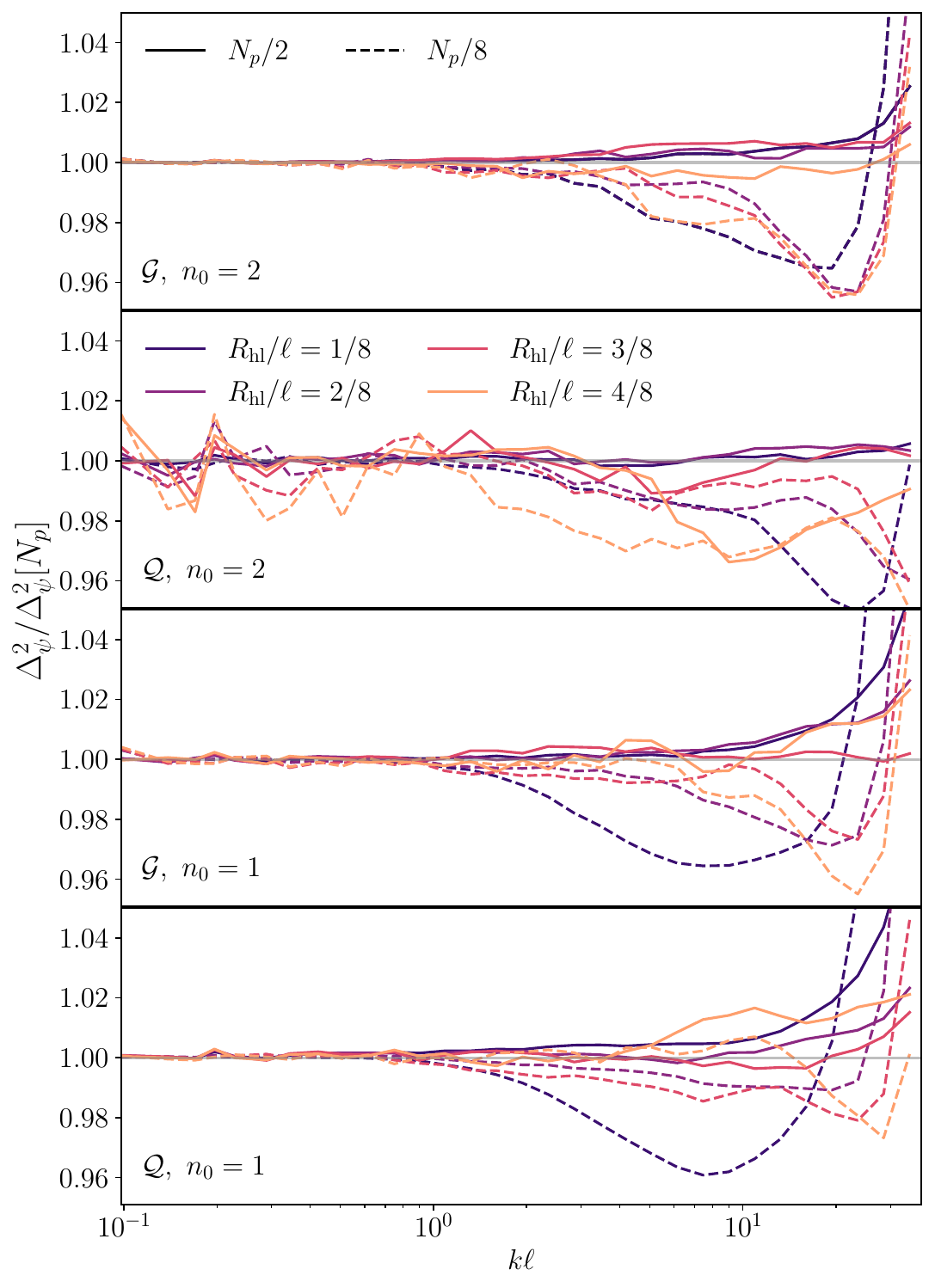}
        \caption{Similar to Fig.~\ref{fig:xiz_np} except showing the power spectrum.  Large scales appear well converged to much better than a percent, except for the $\mathcal{Q}$ case with $n_0=2$ where it is at most around $2\%$.}
        \label{fig:pk_np}
    \end{figure}

    On scales $r\gtrsim\ell$, we see a substantial amount of noise in the ratio.  However, we suspect this is related to the calculation method, even though it doesn't appear as much in the scale-free case with $\mathcal{G}$ and $n_0=2$.  Fig.~\ref{fig:pk_np} shows the analogous power spectrum.  Here convergence on large scales ($k\ell\ll1$) appears much better with sub-percent differences in all cases except the $\mathcal{Q}$ simulation with $n_0=2$.  For the latter case, we do find noise at the level of a couple percent.  We suspect this may be due to the force on large scales being substantially suppressed, although as before it must be in conjunction with the longer runtime as the noise is not visible for the $n_0=1$ case.  Overall, however, the differences in Fig.~\ref{fig:xiz_soft}, \ref{fig:xiz_np} and \ref{fig:pk_np} are generally much smaller on scales $10^{-1}\ell\lesssim r \lesssim \ell$) compared to the $\sim5\times$ and $\sim1.5\times$ enhancement we see in the bottom panels of Fig.~\ref{fig:xiz_n2} and \ref{fig:xiz_n1}.  

    Lastly, we note a technical issue present in the simulations where the largest discrete Fourier mode was incorrectly aliased to the opposite sign.  The only practical effect is a change in the floating point representation of zero, i.e.~the difference in representation between $\sin(\pi)$ and $\sin(-\pi)$.  We tested this explicitly and found that the resulting changes due to floating-point arithmetic are consistent with setting the value to $0$ exactly.  We conclude then that our results are robust to numerical parameters.

\end{document}